\documentclass[letter]{aa}
\usepackage[varg]{txfonts}
\usepackage{graphicx}
\usepackage{natbib}
\bibpunct{(}{)}{;}{a}{}{,} 

\begin{document}

\title{Updated sunspot group number reconstruction for 1749--1996 using the active day fraction method}

\author{T. Willamo\inst{1}
\and I.G. Usoskin\inst{2,3}
\and G. A. Kovaltsov\inst{4}
}

\institute{Department of Physics, University of Helsinki, Finland
\and Space Climate Research Unit, University of Oulu, Finland
\and Sodankyl\"a Geophysical Observatory, University of Oulu, Finland
\and Ioffe Physical-Technical Institute, 194021 St. Petersburg, Russia
}

\date{}

\abstract {}
{Sunspot number series are composed from observations of hundreds of different observers that requires
 careful normalization of the observers to the standard conditions.
Here we present a new normalized series of the number of sunspot groups for the period 1749--1996.
}
{The reconstruction is based on the active day fraction (ADF) method, which is slightly updated with respect to the previous works,
 and a revised database of sunspot group observations.}
{Stability of some key solar observers has been evaluated against the composite series.
The Royal Greenwich Observatory dataset appears fairly stable since the 1890s but is about 10\% too low before that.
A declining trend of 10--15\% in the quality of Wolfer's observation is found between the 1880s and 1920s,
 suggesting that using him as the reference observer may lead to additional uncertainties.
Wolf (small telescope) appears fairly stable between the 1860s and 1890s, without any obvious trend.
The new reconstruction reflects the centennial variability of solar activity as evaluated using the singular spectrum
 analysis method.
It depicts a highly significant feature of the Modern grand maximum of solar activity in the second half
 of the 20th century, being a factor 1.33--1.77 higher than during the 18--19th centuries.}
{The new series of the sunspot group numbers with monthly and annual resolution, available also in the electronic format,
 is provided forming a basis for new studies of
 the solar variability and solar dynamo for the last 250 years.
}

\maketitle

\section{Introduction}

Sunspots, the dark spots on the Sun, are easy to observe even with a basic optical instrument, and this was done
 by many generations of professional and amateur astronomers throughout the centuries.
As a result, the sunspot number forms the longest systematic scientific series used as a quantitative index
 of the level of variable solar activity \citep{hathawayLR}.
Because of its length, the sunspot number series includes records from hundreds of observers with different
 optical instruments, measuring/recording techniques, habits, etc.
This unavoidably calls for a need to reduce the data of individual observers to a `standard observer',
 which implies not only a person but material/instrument and conditions.
Because of this, the sunspot number is a relative number.

The first consistent sunspot number series was produced by Rudolf Wolf of Z\"urich who calibrated different observers
 to his own observational conditions.
To reduce data from different observers, Wolf used a simple linear scaling of sunspot counts from each observer
 to standard observers (the so-called $k-$factors).
This is often referred to as daisy-chaining, especially when the number of standard observers is large.
Later, Wolf sunspot number (WSN) series was continued as the international sunspot number series (ISN)
 at the Royal Observatory of Belgium and the Solar Influences Data Center, Sunspot Index and Long-term Solar
 Observations (SILSO), http://www.sidc.be/silso/).
However, several inhomogeneities have been found in the WSN/ISN, and an updated ISN series (version 2, denoted
 as ISN\_v2) had been released \citep{clette14}.
It is important to notice that ISN\_v2 uses Adolf Wolfer, not Rudolf Wolf, as a ``standard observer'' leading to
 the higher (by a factor of 1.667) overall ISN values vs. the `classical' WSN/ISN datasets.
The ISN\_v2 still uses the $k-$factor methodology for calibration of different observers.
We also note that the original raw data for the WSN series are not available in a digital format, making
 a full revision of this series impossible now, although a progress in this direction is on its way \citep{friedli16}.

The WSN/ISN series is based on the counts of both sunspot groups and individual sunspots, with the former
 being weighted with a factor of ten:
\begin{equation}
R = k\cdot(10\cdot G + S),
\end{equation}
where $G$ and $S$ are the numbers of sunspot groups and individual sunspots, respectively, and $k$ is a correction factor,
 characterizing each observer.
However, resolving individual spots may be imprecise with poor instrumentations, and a new series, based only on
 sunspot groups was proposed, called the group sunspot number, GSN, accordingly \citep{hoyt94,hoyt98}.
The GSN is more robust regarding observational conditions than WSN \citep[e.g.,][]{usoskin_LR_17}.
There is still a potential problem related to the grouping of individual spots, which might have been considered by earlier
 observers differently from our present knowledge \citep{clette14}.
This uncertainty is related to both WSN/ISN and GSN but can be fixed by redefining groups in historical sunspot drawings \citep{arlt13}.
The GSN series produced by \citet{hoyt98} also uses the linear scaling and daisy-chaining method to reduce different data
 to the same reference observer, for which the Royal Greenwich Observatory (RGO) was chosen.
The GSN is constantly scaled up by a factor 12.08 to make it comparable with the WSN series.
The main advantage of the GSN series is that \citet{hoyt98} had collected and published the original database of
 raw data, including all the records of individual observers.
This makes it possible to revise the entire series if needed.
Since some corrections and additions have been recently made to this dataset, a revised database
 of the sunspot group numbers, separately for each observer, is published \citep{vaquero16}.
It is referred to as V16 hereafter.
The GSN series was revised by \citet{svalgaard16} who performed a full re-calibration of the observers
 using a modified daisy-chaining method with a reduced number of links: it is called the ``backbone'' method.
The revised ``backbone'' GSN series suggests that the level of solar activity was quite high in the 18th and
 19th centuries, much higher than that implied by the original GSN series by \citet{hoyt98} and by WSN.

Thus, all the earlier series were based on the parametric $k-$factor calibration method (daisy-chain, linear scaling).
However, it has been shown recently \citep{lockwood_ApJ_16,lockwood_SP3_2016,usoskin_k_16} that the linear $k-$factor
 methodology may be inaccurate when applied to sunspot numbers and needs to be replaced by a modern non-parametric method.
Two such methods have been proposed recently: the active-day fraction (ADF) method \citep[][, called U16 henceforth]{usoskin_ADF_16} and the
 method based on the ratio of the number of individual sunspots to that of sunspot groups \citep{friedli16}.
Both these methods use absolute calibration of observers to the standard one and are thus free of
 daisy-chaining and arbitrary choices.

Here we provide a new sunspot number reconstruction using the ADF method, originally introduced by \citet{usoskin_ADF_16},
 the revised and corrected dataset of the sunspot groups \citep{vaquero16}, a larger set of observers, and a slight
 refining of the calibration method and estimate of its uncertainties.

\section{Data}

\subsection{The reference dataset}

As the reference dataset we used, similarly to U16, the database \footnote{available at http://solarscience.msfc.nasa.gov/greenwch.shtml}
 of sunspot groups of the Royal Greenwich Observatory (RGO).
The RGO data is available since 1874, but the early part of the database may suffer from unstable quality.
It is still debated \citep{cliver16} what period may be affected by this (see Section \ref{sec:stability}), but it is conservatively considered that
 the series is fairly homogeneous since 1900 \citep{clette14,usoskin_ADF_16}.
Accordingly we use the RGO data for the period 1900--1976 as the reference dataset to calibrate observers, but the whole RGO dataset
 (1874--1976) is included into this work as an observer (see below).
This period includes seven complete solar cycles, \# 14 through 20.
As the group size we used the observed (uncorrected for foreshortening, viz. as observed from Earth) umbral area of the sunspot groups
 in units of msd (millionths of the solar disk).
We have tested the robustness of the results against the exact length of the reference dataset (Sect.~\ref{sec:RGO}).

\subsection{Observers}

Here we considered major observers with long records of sunspot data covering the periods of the 18th through 20th centuries.
We used the same set of observers as in U16, except of Stark whose reliability is
 unsettled \citep{hoyt98}, and added 11 more observers for the 20th century, extending the database till 1996,
 which is comparable with the GSN \citep{hoyt98}.
As data for individual observers, we used the daily number of sunspot groups collected by \citet{vaquero16}.
This database \footnote{available at http://haso.unex.es/?q=content/data} is based on the initial data
 of sunspot group records gathered by \citet{hoyt98} but includes important updates and corrections.

Data for Schwabe were taken from a recent compilation by \citet{arlt13}
 \footnote{www.aip.de/Members/rarlt/sunspots/schwabe, version 1.3 from 12 August 2015} based on digitized
 drawings and notes of Schwabe.
In this compilation, sunspots were re-grouped using modern definition of sunspot groups which is different
 from the original Schwabe grouping \citep{pavai15}.

\begin{figure*}
\begin{center}
\includegraphics[width=11cm,height=22cm,clip=]{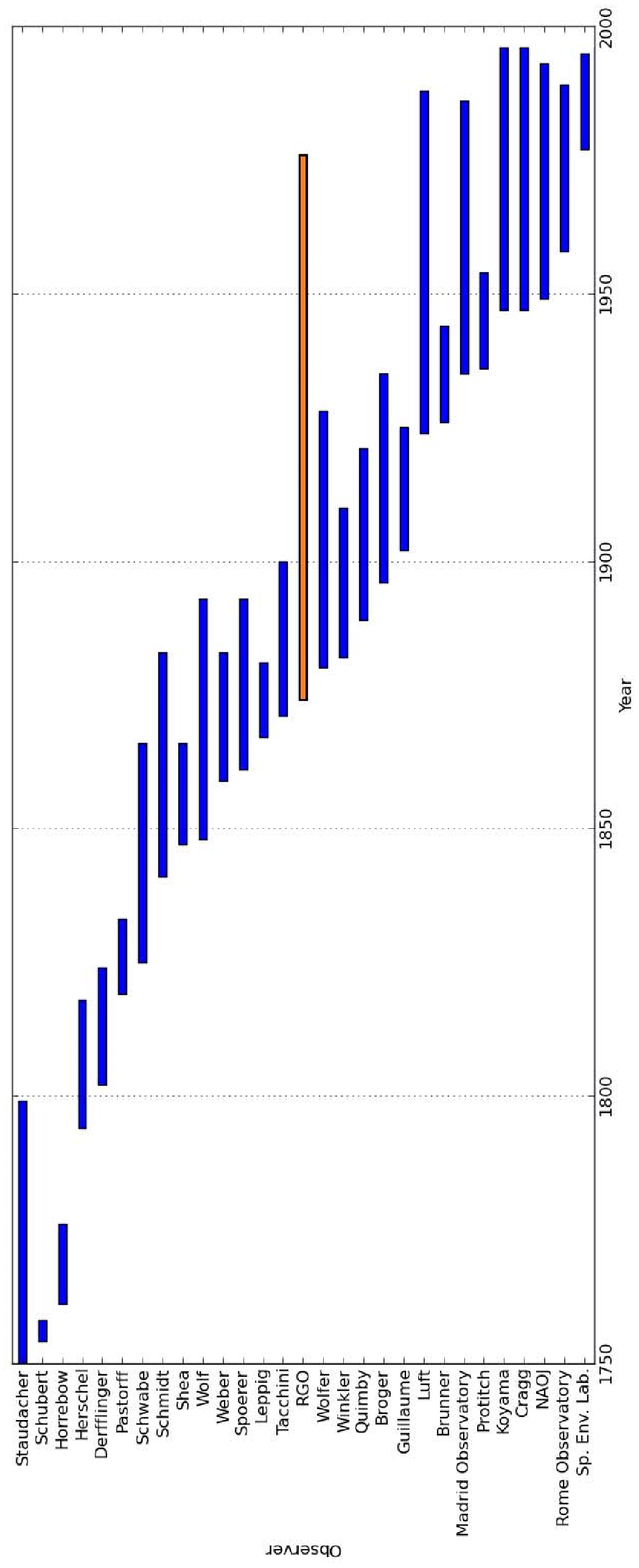}
\end{center}
\caption{The observational periods of the observers used in this study (see also Table~\ref{Tab:observers}).
The reference data set of RGO is shown in orange color.
Note the periods used for calibration of the observers may be shorter than the total observational
 periods shown here.}
\label{fig:coverage}
\end{figure*}

All the observers used in this study are listed in Table~\ref{Tab:observers}.
Their data coverage is shown in Figure~\ref{fig:coverage}.

\section{Calibration method}

\subsection{Calibration of observers}
\label{sec:calib}
Each observer has been calibrated to the reference RGO dataset, following
 the ADF method invented by U16.
The ADF is the ratio of active days (with at least one sunspot group reported) to the total number of
 observational days per month.
The method is based on comparing the ADF statistic of an observer to be calibrated with that of the reference dataset.
The fraction of active days within a month is a robust
 indicator of solar activity around solar minima and makes it possible to calibrate different observers
 \citep{harvey99,kovaltsov04,vaquero12,vaquero15}.
Here we have slightly refined the original method, in particular in the part related to the compilation of monthly values (see Section~\ref{sec:mon}).
We have also revised the indirect calibration of Staudacher (see Section~\ref{sec:stau}).

The ADF method used here is slightly modified with respect to the original one (U16) in the following:
\begin{itemize}
\item
When computing the ADF for individual observers, we did not apply here the limitation of considering only months with the number
 of observational days $n\geq 3$, applied in U16.
This makes almost no effect for observers with sufficiently high observation frequency, in particular in the
 19th--20th centuries, but may distort the statistic for observers with low data coverage and uneven
 distribution of observational days.
Accordingly, we have applied this limitation for Derfflinger and Hershel whose data coverage fraction was 11\% and 5\%,
  respectively.
\item
When constructing the conversion matrix (Section~\ref{sec:matrix}) we accounted for the uncertainties in the definition
 of the observational threshold $S_s$, while only the mean $S_s$ values were used by U16.
\item
Data of Staudacher were calibrated differently here (see Section~\ref{sec:stau}).
\item
When calculating the monthly mean $G-$values and its uncertainties from daily values we used here a Monte Carlo method,
 while a weighted averaging method was used by U16.
\end{itemize}
The effect of these improvements are discussed in Section~\ref{sec:comp}.

\subsection{Assessment of the observer's quality}

The calibration method is based on the idea that the `quality' of each observer is characterized by his/her observational acuity, or
 an observational threshold area $S_S$.
The threshold implies that the observer can see and report all the groups with the area larger than $S_S$, while missing
 all smaller groups.
Here we assume that the observational threshold is constant for an observer during the entire period of his/her observations
 but a time variability of the acuity can be considered in subsequent works.
In Section~\ref{sec:stability} we discuss this issue in more detail.
The reference dataset of RGO is assumed to be `perfect' in the sense that RGO does not miss any spots (viz. $S_S=0$).

Similarly to U16, we first made `calibration' curves using the reference dataset.
As a calibration curve we used the cumulative distribution function (CDF) of the occurrence, in the reference
 dataset, of months with the given ADF.
A family of such curves was produced for different values of $S_S$ (all sunspot groups
 with the area smaller than that were considered as not observed).
Thus, each calibration curve uniquely corresponds to a value of the observational threshold $S_S$.
Calibration curves were produced for different values of the filling factor $f$
 (the ratio of the number of days with reported observations, including no-spot observation, to the total number of
 days during the observation/calibration period), by randomly removing $(1-f)$ fraction of daily values from
 the RGO reference dataset to simulate a realistic observer.
We performed 100 such random sub-samplings and calculated the mean and the asymmetric two-tail 68\% confidence
 intervals for each case.
\begin{figure}
\begin{center}
\includegraphics[width=0.95\columnwidth]{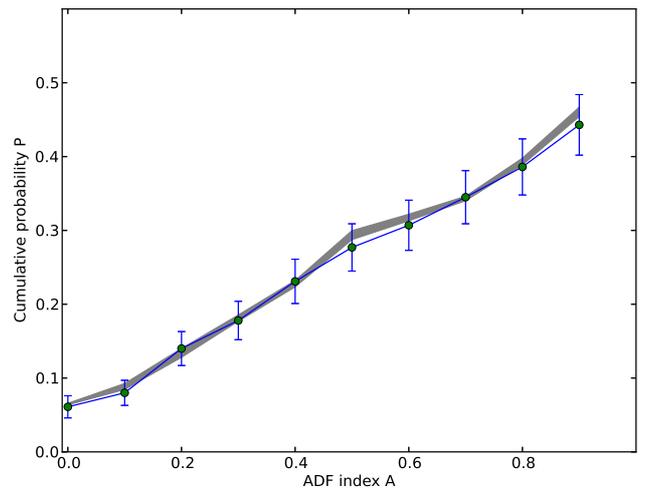}
\caption{The cumulative distribution function of the reference dataset for
 $S_S=25$ msd and the filling factor $f=0.92$ (grey line with $1\sigma$ uncertainties)
 compared to that for Quimby (dots with error bars).
}
\end{center}
\label{Fig:cal}
\end{figure}

For each observer we constructed a CDF curve using his/her observations during the calibration period.
The ADF and the subsequent CDF were calculated for all months with observations, not only months
 with three or more observational days as was done in U16.
 This limitation was, however, applied to Derfflinger and Hershel.
An example of the CDF curve for observer Quimby is shown in Figure~\ref{Fig:cal}.
It is important that solar activity during the calibration period roughly corresponds to that in the reference
 dataset U16.
The reference dataset covers a wide range of solar cycles, from moderate cycle 14 to the very high cycle 19,
 but weak cycles are not presented there, thus this method cannot work reliably for the periods of grand minima
 such as the Maunder or Dalton minimum.
Accordingly, we selected for calibration of each observer periods with a relatively good coverage by the data
 and covering full solar cycles (except for the case of Schubert -- see U16).
If the observer had a sufficiently long period of direct overlap with the reference dataset,
 the period of overlap was used for calibration.
In these cases the reference calibration curves were also calculated for the same overlap period.
The list of the selected observers and their calibration periods is presented in Table~\ref{Tab:observers}.

The observational threshold for each observer was defined by fitting the family of the calibration curves to the
 actual CDF curve of this observer, as shown in Figure~\ref{Fig:cal}.
The best-fit value of $S_S$ and its 68\% ($\pm 1\sigma$) confidence interval were defined by the $\chi^2$ method.
The minimum value $\chi_0^2$ corresponds to the best-fit estimate of the observational threshold, while the values
 of $S_S$ corresponding to $\chi_0^2+1$ bound the 68\% confidence interval.

The values of the acuity observational threshold $S_S$ are shown along with the 68\% confidence intervals
 in the last column of Table~\ref{Tab:observers} for each observer.
One can see that it varies from very small numbers around zero for good observers up to 60--70 msd for
 poorer observers.
In the cases of the National Astronomical Observatory of Japan (NAOJ) and Space Environment Laboratory (SEL),
 we found that their quality is better than that of RGO, i.e. a formally negative threshold would have been obtained
 in the calibration.
Since the negative threshold cannot be defined for the reference dataset, we further consider no threshold for them,
 assuming them to be on the same observational quality as RGO.
The negative threshold would lead to a slight overestimate (1--2\%) of the
 values of the final $G-$series during the second half of the 20th century.

\subsection{Correction matrix}
\label{sec:matrix}

\begin{figure}
\includegraphics[width=\columnwidth]{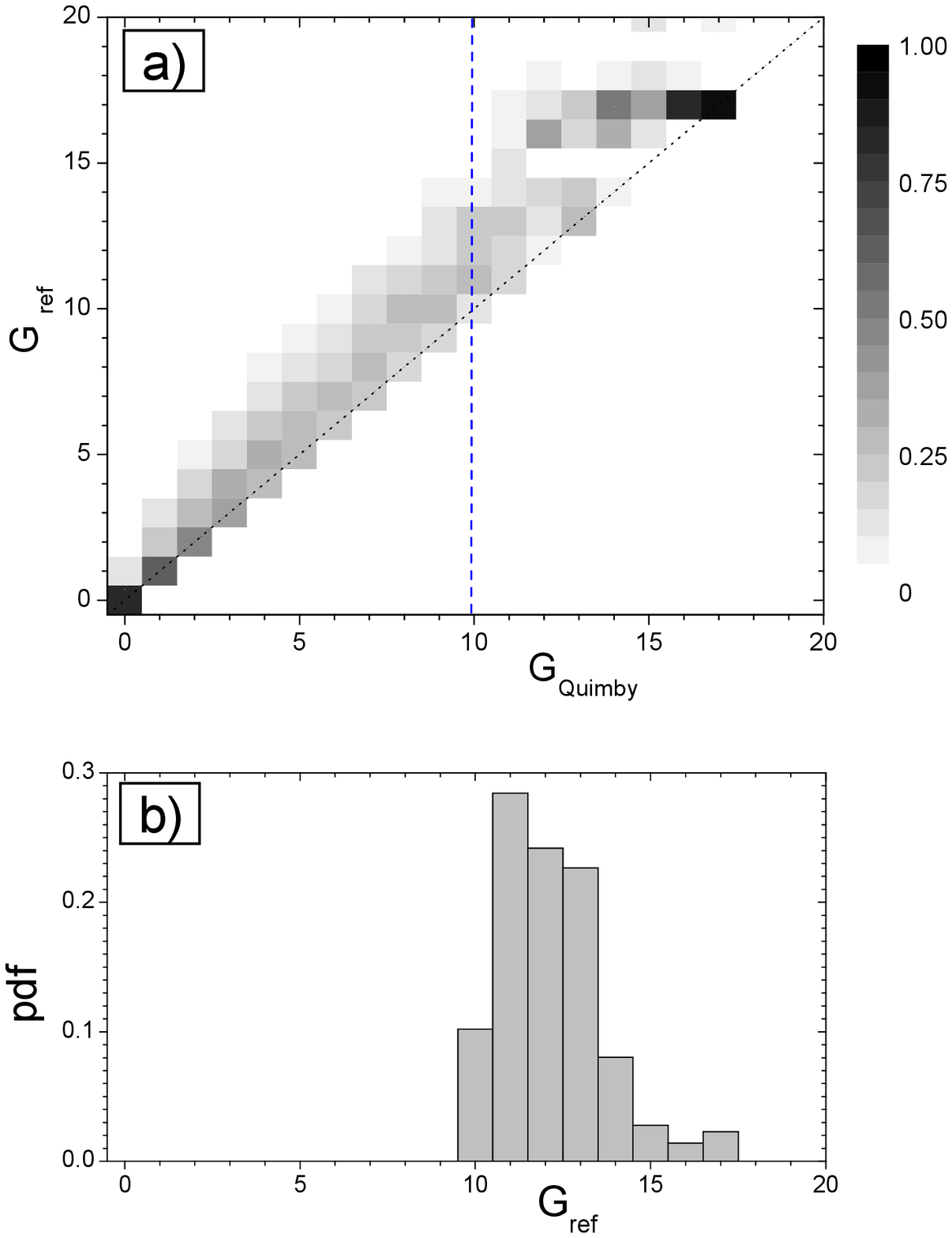}
\caption{Panel a): Correction matrix for Quimby giving the probability to find the value of $G_{\rm ref}$ in the reference dataset
 on a day when $G_{\rm Quimby}$ sunspot groups were reported by Quimby.
 Panel b depicts a pdf (probability density function) of $G_{\rm ref}$ to be found in the reference dataset for days
  whith $G_{\rm Quimby}=10$ (the cross-section of panel a at the blue dashed line).  }
\label{Fig:matrix}
\end{figure}

Once the observational threshold $S_S$ has been defined for an observer, a correction matrix can be constructed
 in the following way.
From the entire reference dataset, a distribution of the daily values of $G_{\rm ref}$ (the number of sunspot groups of all
 sizes in the reference dataset for a given day) as function of $G_{S}$ (the number of sunspot groups with the size $\geq S_s$
 for the same day) is constructed and normalized to unity in the `vertical' direction so that it gives a probability
 to observe the `true' number of groups $G_{\rm ref}$ for a day when the observer reported $G_{S}$ groups.
In order to account for the uncertainties of the defined value $S_S$ the matrix was constructed not only for the
 best-fit values (as done by U16) but averaging matrices for all the (integer) values of $S_S$ from the corresponding
 68\% confidence interval.
By construction, $G_{\rm ref} \geq G_S$.
An example of the correction matrix is shown in Figure~\ref{Fig:matrix} for Quimby.

Such matrices are further used to correct the actual observation with a given threshold value to the reference level.

\subsection{Calibration of Staudacher}
\label{sec:stau}

The only observer who cannot be calibrated directly using the ADF methods is Staudacher since he did
 not report spotless days properly (see U16 for more detail).
On the other hand, Staudacher was a key observer covering the second half of the 18th century
 (although with sparse observations), and it is important to consider him for that period.
Since the data from Staudacher overlaps with observations of two other observers, Horrebow and Schubert,
 who can be calibrated using the ADF method, we have performed an indirect correction of the
 Staudacher data via Horrebow and Schubert.

For all the days with reported observations of Staudacher, we checked Schubert's and Horrebow's data for
 observations on the same day.
If none was found, we checked for observations on the previous day and, if none was found, on the next day.
If no observations of Schubert or Horrebow were found on the neighboring days, we checked for the available
 data two days before, and finally two days after the day with Staudacher's observation.
We have checked that each Staudacher's observation was used not more than once in the analysis.
Although using the observations from 1--2 neighboring days may introduce a small uncertainty due to short-living
 small groups \citep[e.g.][]{willis16}, this is outweighed by the improvement of statistic.
We found 138 days, when observations of Staudacher coincided with data from either Schubert or Horrebow,
 120 days when the observations were separated by one day, and 44 cases when they were separated by two days,
 leading altogether to 302 days for calibration.

\begin{figure}
\includegraphics[width=\columnwidth]{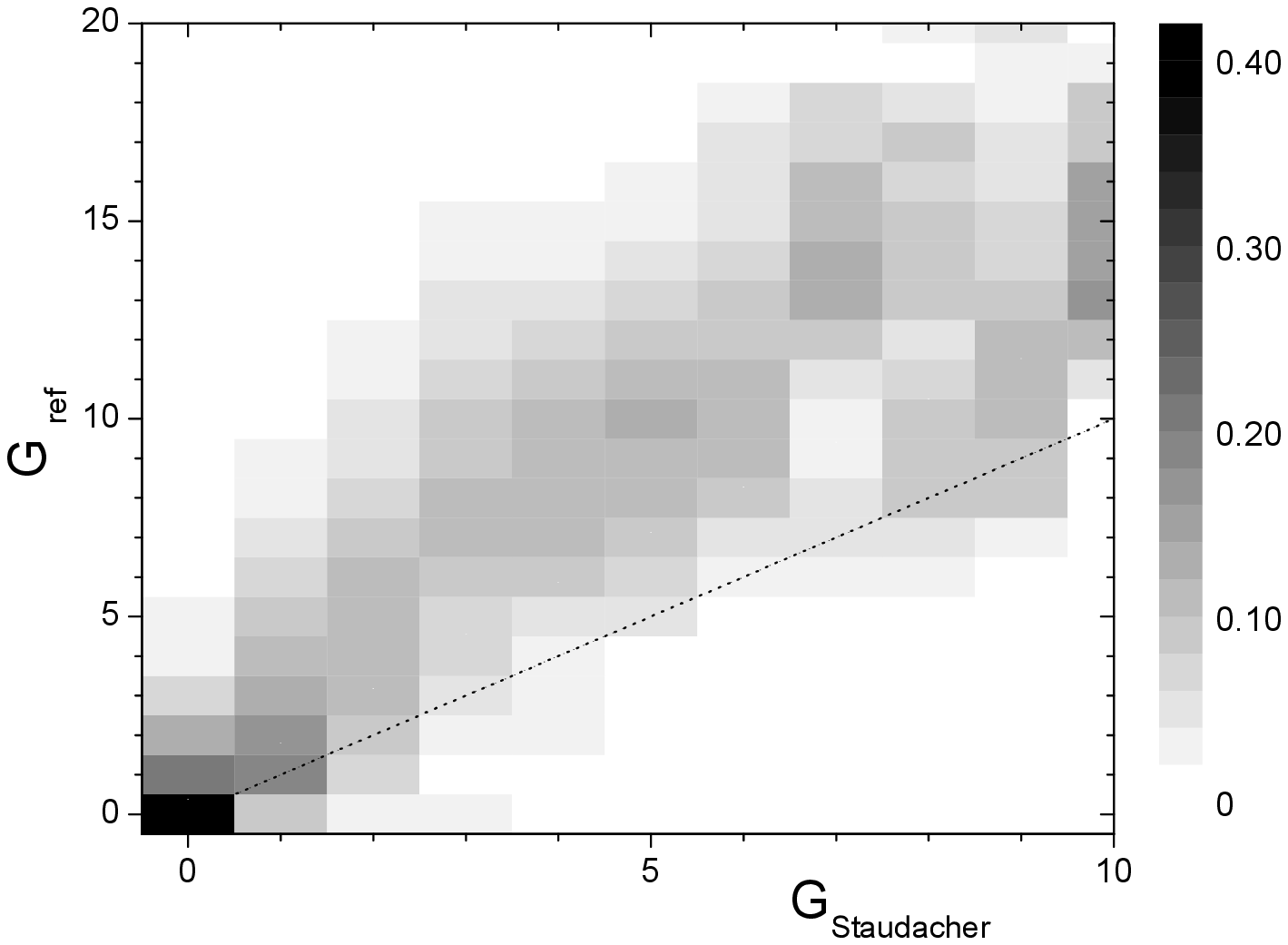}
\caption{Correction matrix of Staudacher giving the probability to find the value of $G_{\rm ref}$ in the reference dataset
 on a day when $G_{\rm Staudacher}$ groups were reported by Staudacher.}
\label{Fig:matrix_S}
\end{figure}

Next, for all daily values of Staudacher $G\mathrm{_{Stau}}$ from the sub-sample described above we collected
 the corresponding daily values of $G^\ast$ for Schubert or Horrebow, already normalized to the
 reference level using the ADF method.
As a result we composed a correction matrix (shown in Figure~\ref{Fig:matrix_S}) which allows us to convert
 the number of sunspot groups reported by Staudacher to the reference observer, in the same way as used
 for other observers.

\begin{table*}
\caption{List of observers used in this work. The columns are:
the name of the observer; observer's ID number \# in the V16 database; total period of observations $T\mathrm{_{obs}}$
 (as shown in Figure~\ref{fig:coverage});
 period used for calibration $T\mathrm{_{cal}}$; the number of observational days during the calibration period $N$;
 the filling factor $f$ during the calibration period; the fraction of active days, ADF, during the calibration period;
 estimate of the observational threshold $S_S$ (in msd) along with the 68\% confidence interval.}
\begin{tabular}{l| c c c c c c c}
\hline
Observer & \# & $T_{\rm obs}$ & $T_{\rm cal}$ & $N$ & $f$(\%) & ADF(\%) & $S_S$(msd) \\
\hline
RGO$^\ast$ & 332 & 1874--1976 & 1900--1976 & 28124 & 100  & 86  & 0 \\
SEL$^\dagger$ & 459 & 1977--1995 & 1977--1995 & 6922 & 100  & 94  & $<0$ \\
Rome Obs.$^\dagger$ & 454 & 1958--1989 & 1958--1976 & 4758 & 69  & 89  & $22(_{18}^{26}) $\\
NAOJ$^\dagger$ & 447 & 1949--1993 & 1954--1976 & 6243 & 74  & 89  & $<0$ \\
Cragg$^\dagger$ & 736 & 1947--2009 & 1954--1976 & 7015 & 84  & 87  & $20(_{17}^{22}) $\\
Koyama & 445 & 1947--1996 & 1953--1976 & 4727 & 54  & 88  & $3(_0^6) $\\
Protitch$^\dagger$ & 438 & 1936--1954 & 1936--1954 & 3357 & 48  & 91  & $0(_0^2) $\\
Madrid Obs.$^\dagger$ & 435 & 1935--1986 & 1935--1976 & 10049 & 66  & 86  & $28(_{24}^{32}) $\\
Brunner$^\dagger$ & 428 & 1926--1944 & 1926--1944 & 4901 & 71  & 88  & $2(_0^6) $\\
Luft$^\dagger$ & 464 & 1924--1988 & 1924--1976 & 7536 & 39  & 87  & $15(_{12}^{17}) $\\
Guillaume$^\dagger$ & 386 & 1902--1925 & 1902--1925 & 6340 & 72  & 79  & $5(_1^{11}) $\\
Broger$^\dagger$ & 370 & 1896--1935 & 1900--1935 & 8600 & 65  & 78  & $8(_5^{11}) $\\
Quimby & 352 & 1889--1921 & 1900--1921 & 7428 & 92  & 73 & $23(_{17}^{31})$ \\
Winkler & 341 & 1882--1910 & 1889--1910 & 4813 & 60  & 75  & $60(_{51}^{71})$ \\
Wolfer & 338 & 1880--1928 & 1900--1928 & 7165 & 68  & 77 & $6(_1^{11})$ \\
Tacchini & 328 & 1871--1900 & 1879--1900 & 6256 & 78  & 82  & $18(_{13}^{22})$ \\
Leppig & 324 & 1867--1881 & 1867--1880 & 2463& 48  & 73  & $50(_{43}^{61})$ \\
Spoerer & 318 & 1861--1893 & 1865--1893 & 5409 & 51  & 86  & $0(_0^2) $\\
Weber & 311 & 1859--1883 & 1859--1883 & 6983 & 76  & 81  & $25(_{20}^{31}) $\\
Wolf & 298 & 1848--1893 & 1860--1893 & 8122 & 65  & 77  & $45(_{36}^{49}) $\\
Shea & 295 & 1847--1866 & 1847--1866 & 5538 & 76  & 80  & $25(_{20}^{31}) $\\
Schmidt & 292 & 1841--1883 & 1841--1883 & 6970 & 44  & 79  & $10(_7^{12}) $\\
Schwabe & 279 & 1825--1866 & 1832--1866 & 8297 & 65  & 86  & $8(_4^{12}) $\\
Pastorff & 263 & 1819--1833 & 1824--1833 & 1462 & 40  & 87  & $3(_0^9) $\\
Derfflinger & 246 & 1802--1824 & 1816--1824 & 374 & 11  & 69  & $50(_{40}^{80}) $\\
Herschel & 236 & 1794--1818 & 1795--1810 & 372 & 5  & 84  & $20(_{10}^{40}) $\\
Horrebow & 180 & 1761--1776 & 1766--1776 & 1365 & 34  & 74  & $70(_{54}^{87}) $\\
Schubert & 178 & 1754--1758 & 1754--1757 & 446 & 31  & 59  & $20(_{14}^{23}) $\\
Staudacher$^a$ & 466 & 1749--1799 & -- & -- & 10  & --  & -- \\
\hline
\end{tabular}
\label{Tab:observers}
\\ Notes:\\
$^\ast$ -- reference dataset\\
$^\dagger$ -- new with respect to \citet{usoskin_ADF_16}; \\
$^a$ -- calibrated indirectly (see Section~\ref{sec:stau}).\\
\end{table*}

\section{Construction of the composite series} \label{sec:mon}

\subsection{Daily values}

\begin{figure}
\includegraphics[width=\columnwidth]{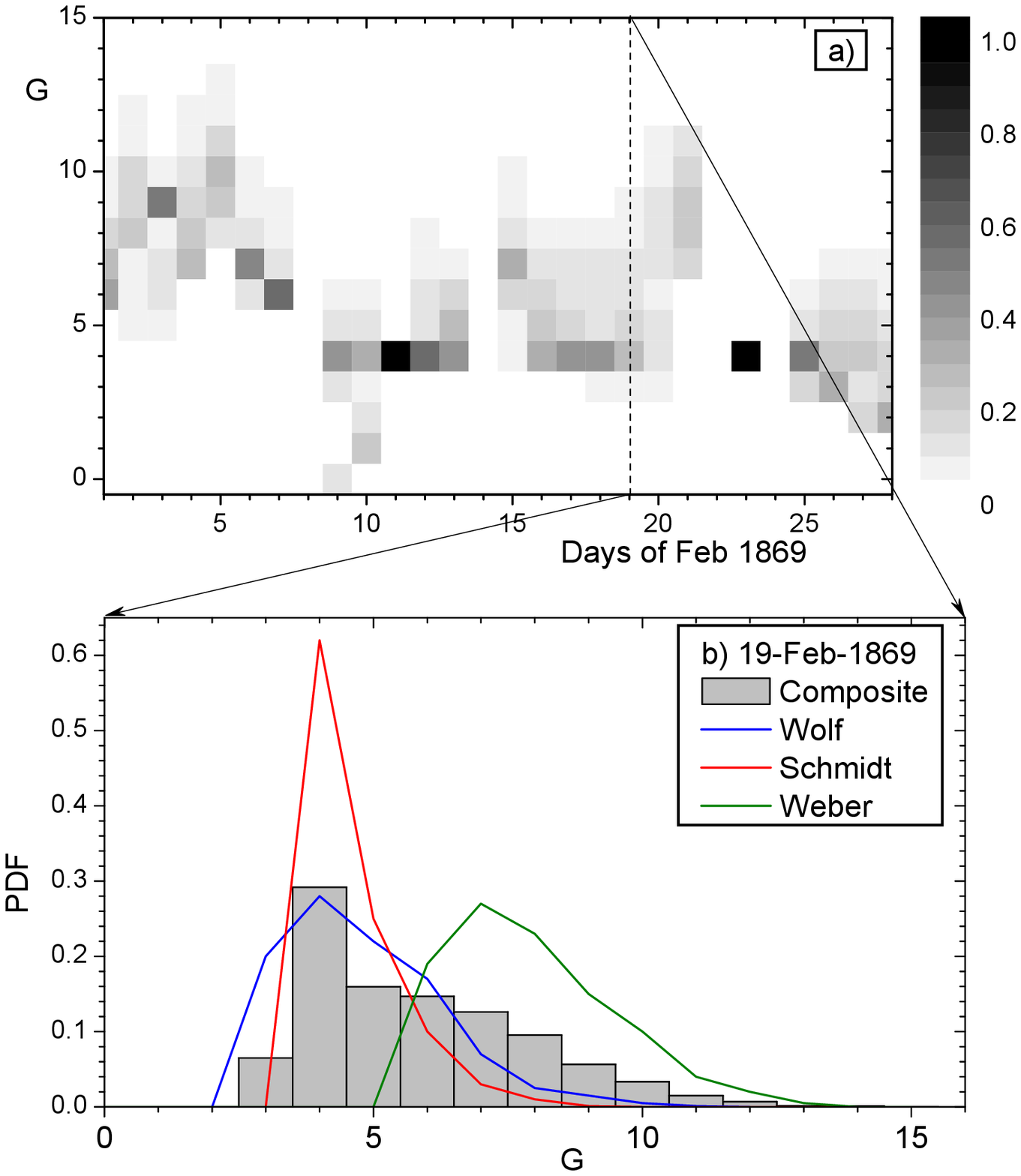}
\caption{Panel a: Distribution of PDF (grey scale on the right) for corrected number of sunspot groups for February 1869.
Panel b: The PDF of the corrected $G$ values for the day of of 19-Feb-1869 (cross-section of panel A at the dashed vertical line).
The composite PDF is shown by grey bars, while PDF distributions for Wolf, Schmidt and Weber are shown as blue, red and green
 lines, respectively.}
\label{fig:daily}
\end{figure}

Using the correction matrices we first calculated PDFs of the corrected (to the reference observer) $G-$values
 for each observer and day.
An example of such PDFs is shown in Figure~\ref{fig:daily}b for the day of 19-Feb-1869
 for three observers whose records are available for this day: Wolf, Schmidt and Weber (colored lines in the Figure).
Next, we made a sum of all the available individual PDFs for the day and renormalized it again to the unity.
This makes a composite PDF of all the observers for this day (grey bars in Figure~\ref{fig:daily}b).
Such composite daily PDFs of the corrected $G-$values were made for all days (see an example for the month
 of February 1869 in Figure~\ref{fig:daily}a).
This dataset (available from the authors upon request) makes a basis for further computations of the monthly time series.

\subsection{Monthly series}
\label{S:m}
Using the daily PDF series discussed above, we constructed the monthly corrected number of sunspot groups using a Monte Carlo
 method.
Within each month we considered all days with available observations.
For each such day we randomly took $G-$value corresponding to the PDF distribution (an example is shown in Figure~\ref{fig:daily}b),
 and then computed the corresponding monthly mean $G-$value as the arithmetic mean.
This procedure was repeated 1000 times so that an ensemble of 1000 monthly values $G$ was obtained.
From this ensemble we calculated the mean and the bounds of the (asymmetric) 68\% two-side confidence interval (corresponding to the generally
 asymmetric $\pm 1\sigma$ interval) for the monthly $G-$value.
For the example shown in Figure~\ref{fig:daily}a (February 1869) the monthly number of sunspot groups was found to be
 $6.10^{+0.37}_{-0.36}$.

There is an uncertainty related to calculation of monthly values from a small number of sparse daily observations,
 when a simple arithmetic average tends to overestimate (by up to 20--25\%) the number of sunspot groups for periods of high activity
 if the number of daily observations per month is smaller than three \citep{usoskin_SP_daily03}.
This may affect the values for the 18th century and Dalton minimum where data coverage was low, giving the
 numbers presented here as a conservative upper limit.
However, this effect does not influence the calibration and correction procedure since they operate with daily data.

\begin{figure}
\includegraphics[width=\columnwidth,clip=]{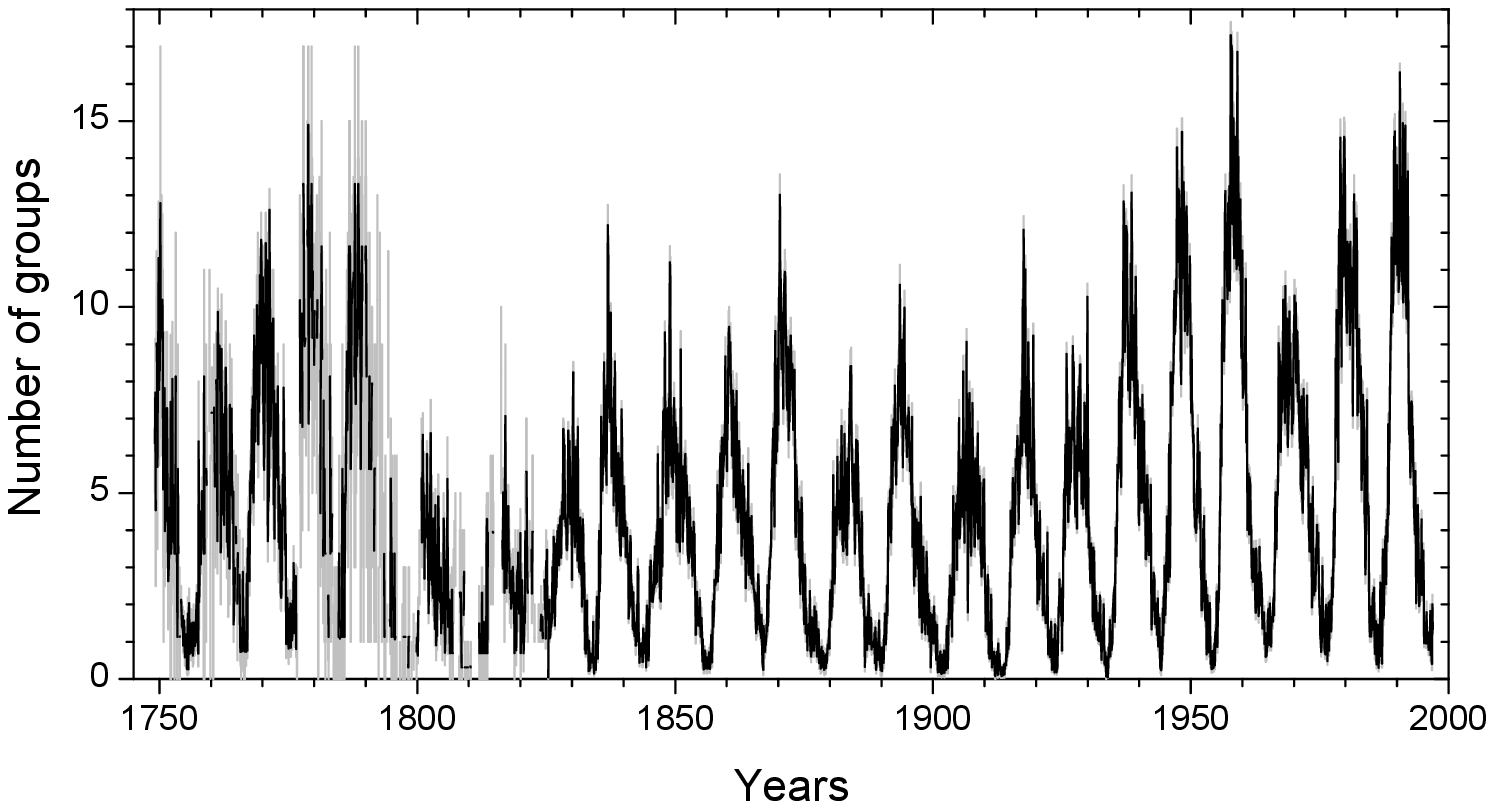}
\caption{Monthly values of the final composite series of number of sunspot groups.
  Error bars (68\% two-side confidence intervals) are shown in grey.
  This series is available in electronic format at CDS.}
\label{fig:monthly}
\end{figure}

The final composite series is shown in Figure~\ref{fig:monthly} and is available in digital format at CDS.

\subsection{Annual series}

\begin{figure}
\centering
\includegraphics[width=\columnwidth,clip=]{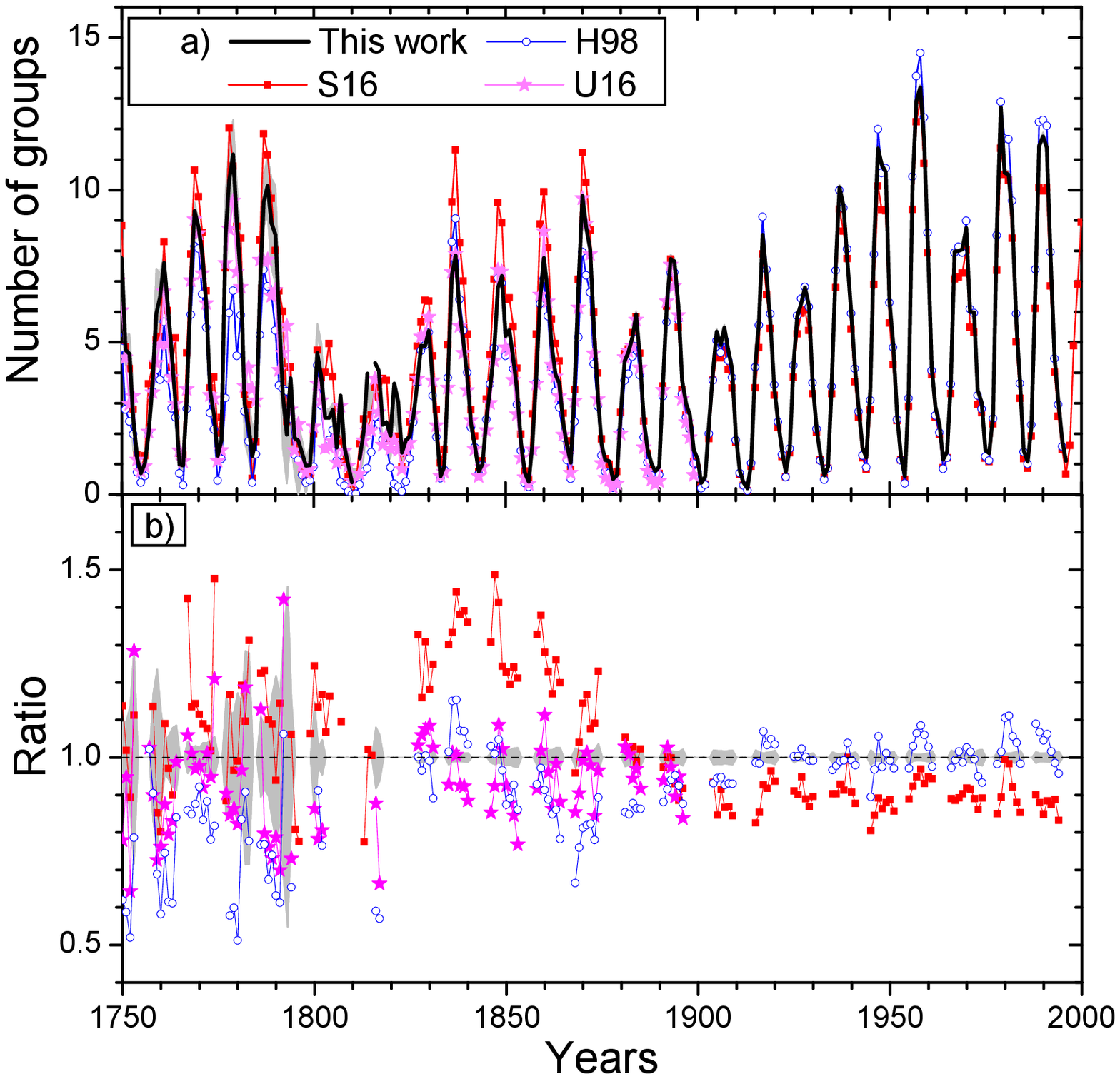}
\caption{Panel a: Annual mean numbers of sunspot groups.
The black line with grey shading depicts the result of this work with the 68\% confidence
 interval.
 Numerical values are given in Table~\ref{Tab:annual}.
 Other colored curves with symbols show reconstructions of $G$ by H98 \citep{hoyt98}, S16 \citep{svalgaard16}
  and U16 \citep{usoskin_ADF_16}.
 Panel b: The ratio between the colored plots (shown in panel a and following the same notation)
  to the result of this work.
 The ratio is not shown for years with low activity ($G<3$). }
\label{fig:annual}
\end{figure}

From the monthly values we computed the annual mean $G-$values and their uncertainties using the Monte Carlo
 method (with 1000 ensemble members) similar to that applied to compute monthly values from daily ones.
The resultant series of the annual numbers of sunspot groups is shown in Figure~\ref{fig:annual}a
 as the black curve with the 68\% confidence intervals shown in grey, and in Table~\ref{Tab:annual}.

\section{Consistency of the result} \label{sec:stability}

\begin{figure*}
{\includegraphics[width=11.5cm,clip=]{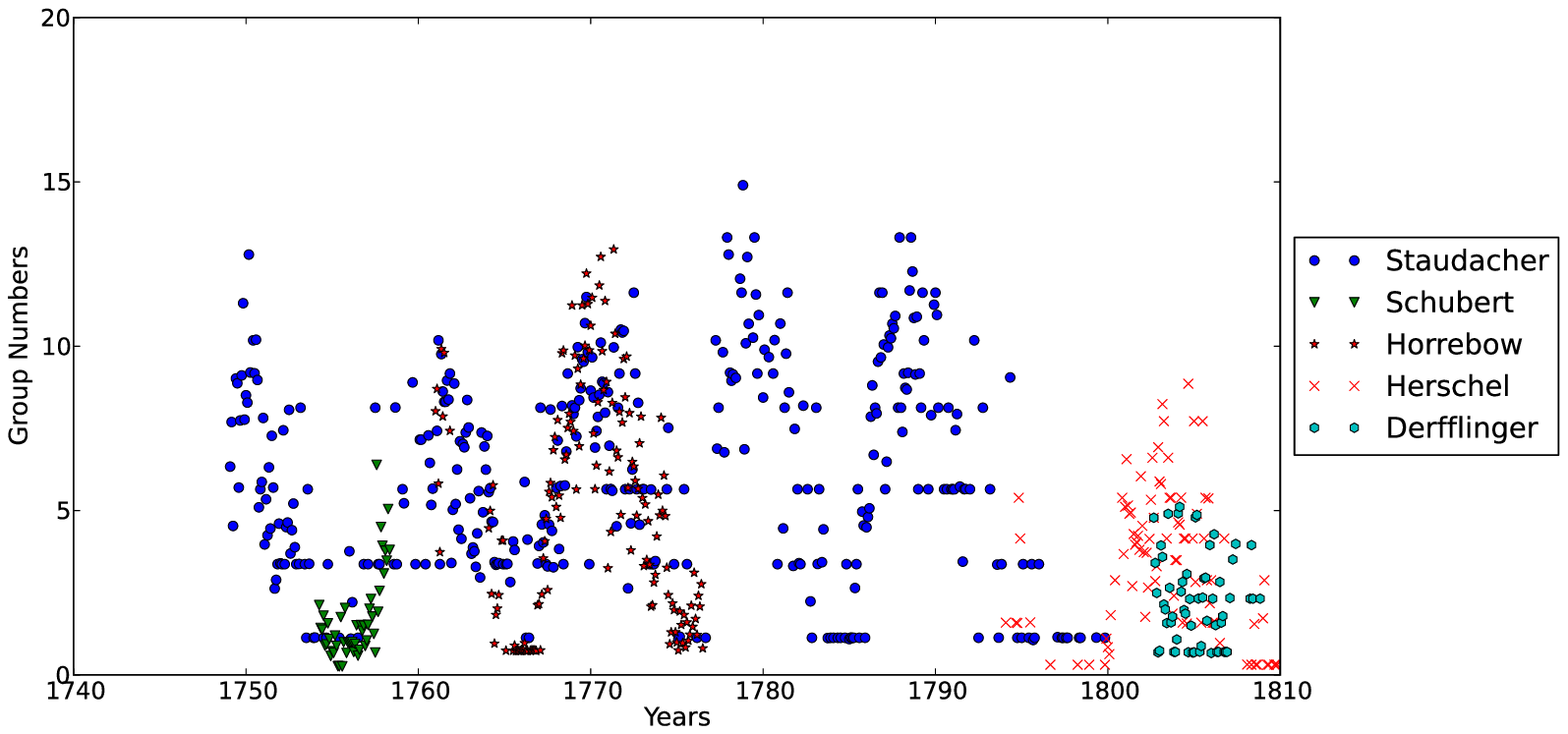}}\\
{\includegraphics[width=11.4cm,clip=]{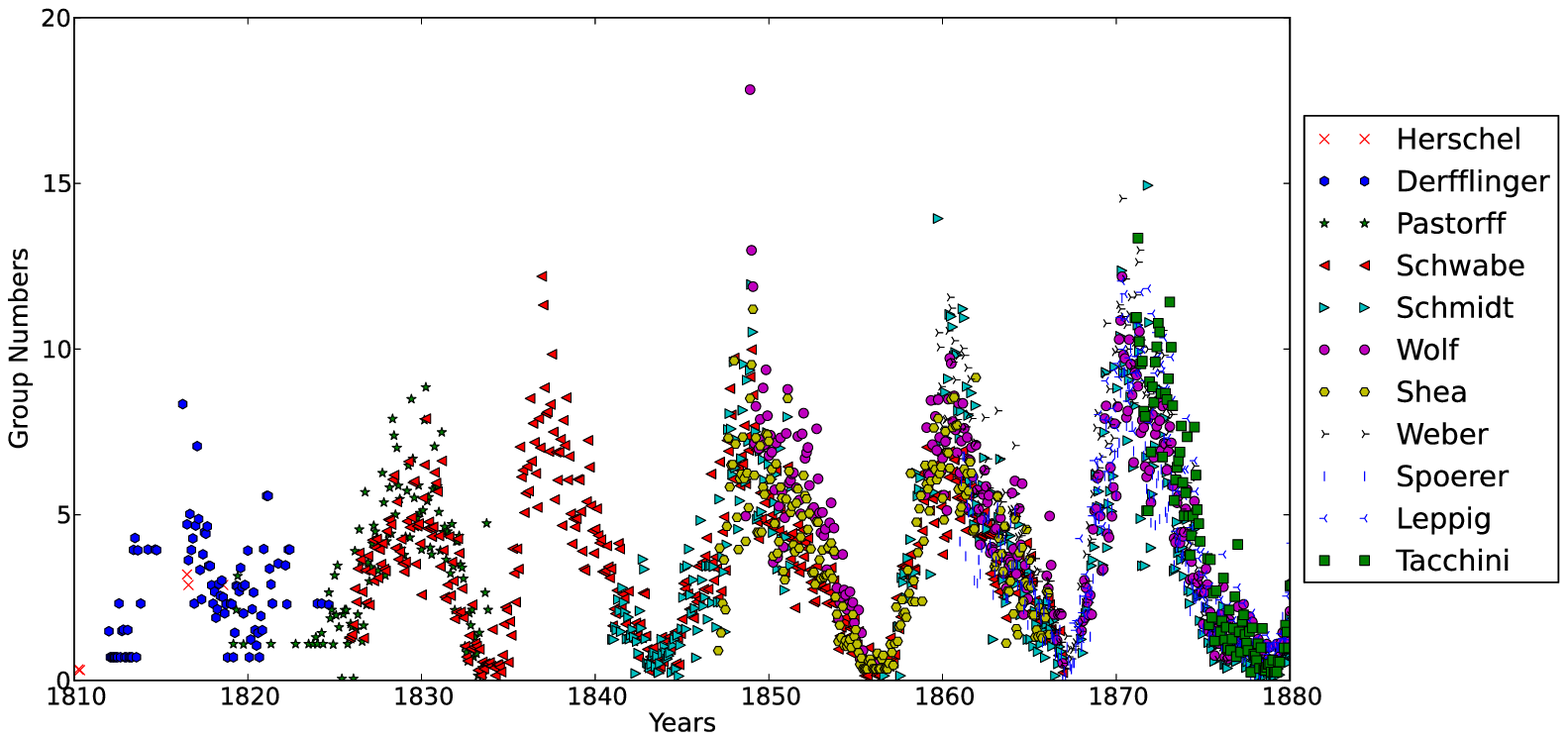}}\\
{\includegraphics[width=11cm,clip=]{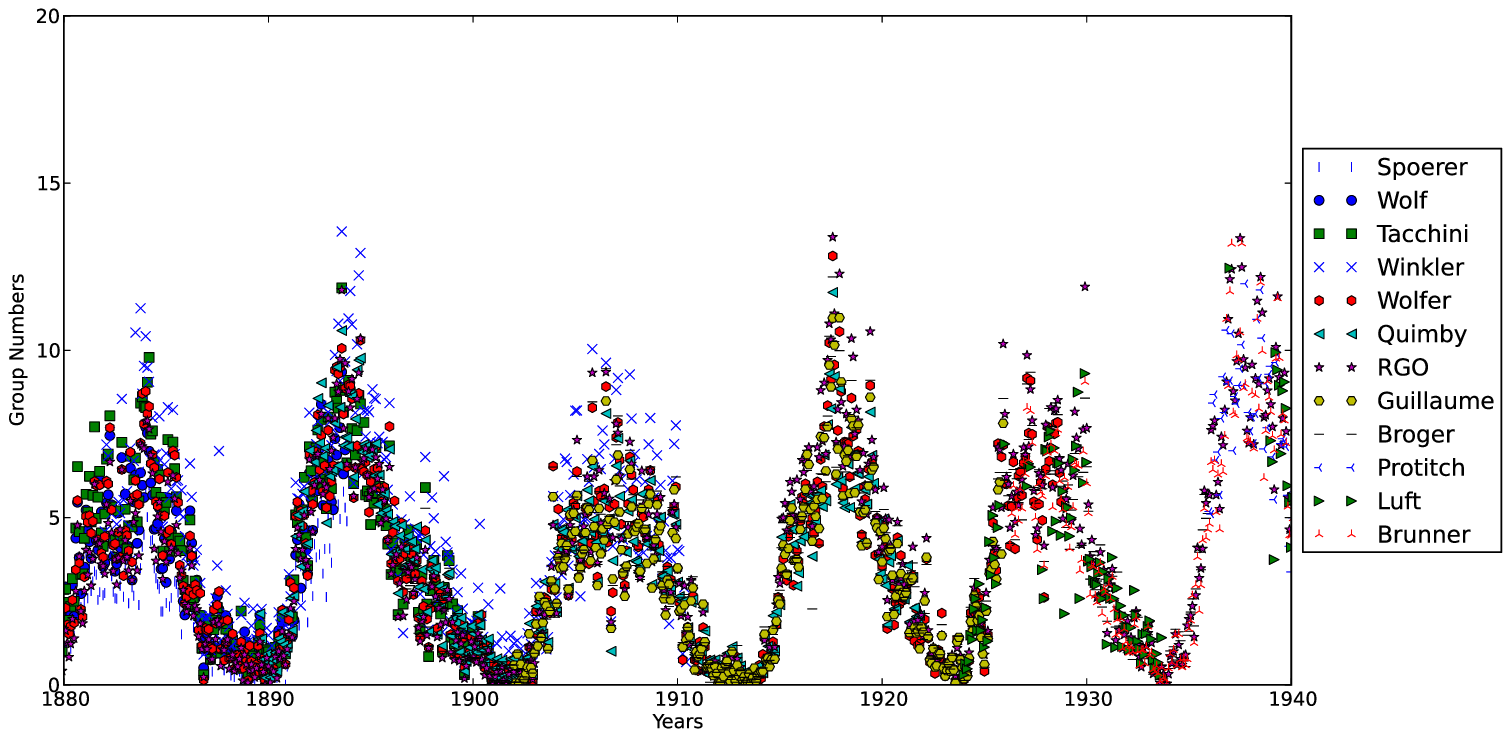}}\\
{\includegraphics[width=12.4cm,clip=]{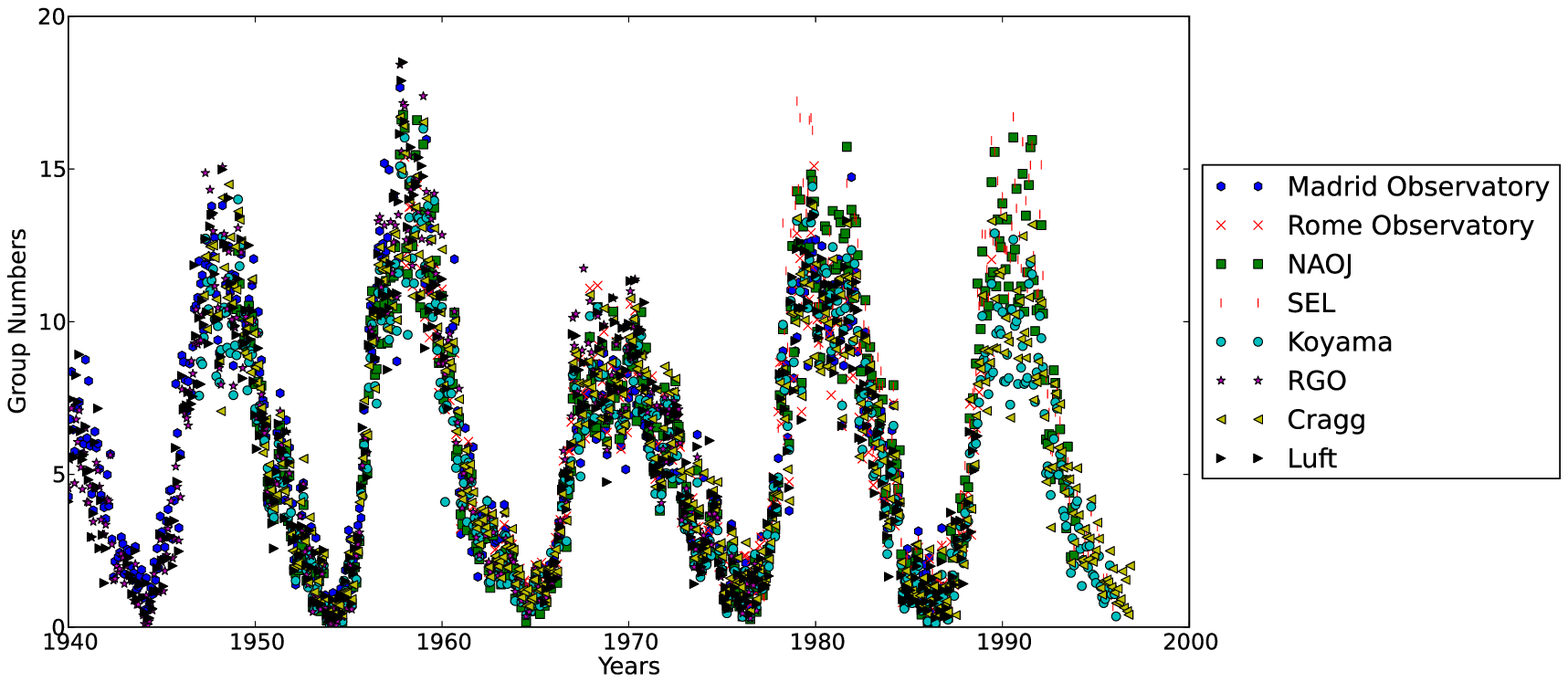}}
\caption{Monthly series of sunspot group numbers by individual observers.}
\label{fig:all-obs}
\end{figure*}

First we computed the monthly series of $G-$values for each observer using the same method as described in Section~\ref{S:m} but
 applying it to data of only this observer (viz. without construction of the composite series).
The resultant series are shown in Figure~\ref{fig:all-obs}.
One can see that there is a good agreement between different observers, especially in the 19th and 20th centuries.
The agreement is worse around the Dalton minimum, when the reconstructions based on data of Herschel and Derfflinger
 diverge, suggesting that the level of solar activity during that period is quite uncertain.
On the other hand we stress that, since the ADF method is free of daisy-chaining and based on a direct calibration
 of the observers to the reference one, the big uncertainty around the Dalton minimum does not affect other periods,
  even before it.

\subsection{Stability of observers}

It is difficult to judge about the stability of observers and their calibration from simply over-plotting
 the series as done in Figure~\ref{fig:all-obs}.
We studied also, as the measure of the observer's stability, the ratio of the $G-$values (annually averaged to avoid noisy data),
 obtained using only data from this observer, to that of the composite series constructed using all but this observer's data.
In order to avoid the ratio of small numbers we excluded years when the mean number of sunspot groups $G$ was below three.

\subsubsection{RGO data}
\label{sec:RGO}
\begin{figure}
\centering
\includegraphics[width=\columnwidth,clip=]{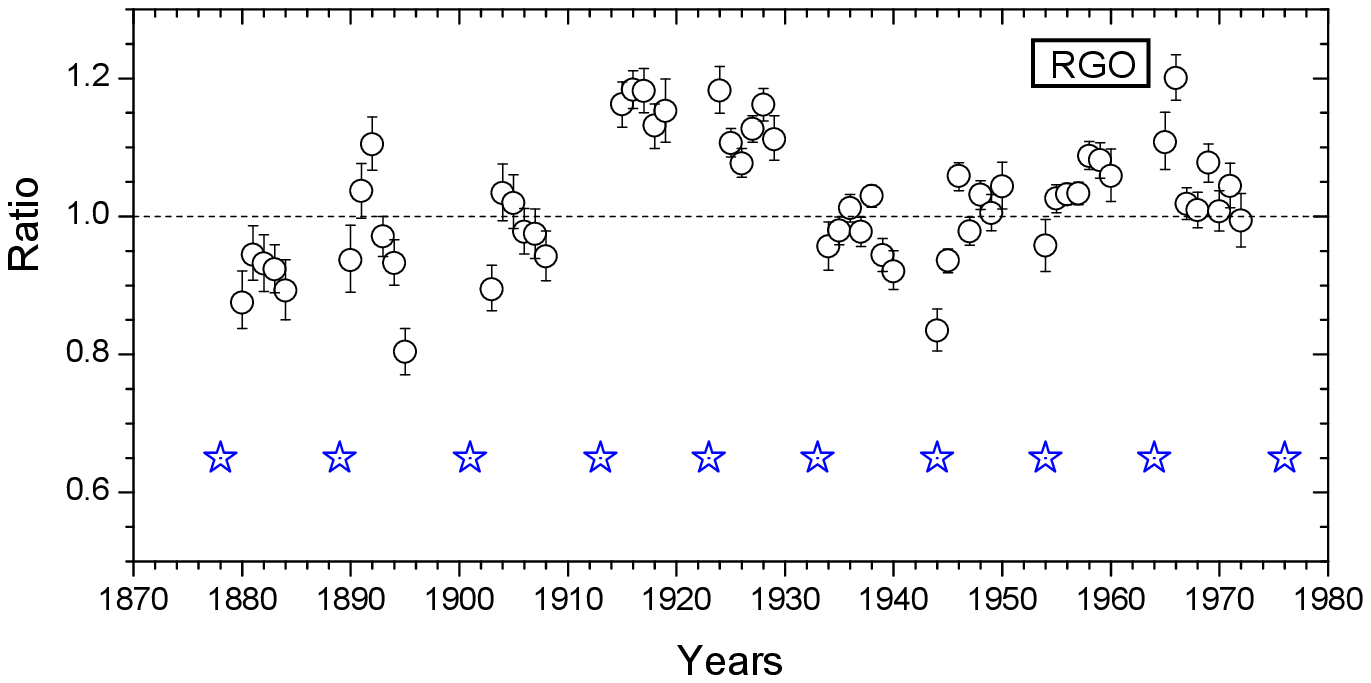}
\caption{The ratio between annual mean $G-$values obtained using only RGO data to those from the composite series
 computed without RGO.
 Ratios for the years with low activity ($G<3$) are not shown.
 Error bars depict the 68\% confidence interval for the ratio.
 Blue stars correspond to the years of official solar cycle minima.}
\label{fig:RGO-ratio}
\end{figure}

As an example, Figure~\ref{fig:RGO-ratio} presents the ratio of the annual $G-$values only from RGO
 to that of the composite series without RGO.
One can see that the ratio is close to unity around solar cycle maxima, as expected if the calibration was done correctly,
 always being within $\pm 20$\%, but there are some systematic features.
The ratio is systematically too low for the first cycle covered by RGO data, before ca. 1890.
This implies that the RGO data underestimates the number of sunspot groups by about 10\% during that period.
This inhomogeneity in the earlier years of the RGO dataset is quite well known \citep[see, e.g.][]{clette14},
 but its exact extent is still debated \citep{cliver16}.
Most studies \citep{sarychev09,carrasco13,aparicio14,willis16} limit the effect of under-counts to the period
 before ca. 1885, which is likely related to the secondary magnifier installed at Greenwich in 1884
 \citep{cliver16}.
However, \citet{cliver16} claim that the inhomogeneity might had extended until ca. 1915.
\citet{clette14} stated that the RGO data is homogeneous at least since 1900.
Our result confirms that the RGO data suffers from the inhomogeneity (10--15\% under-count of sunspot groups)
 only before 1890, while the ratio during the period 1890--1910 is around unity and fully consistent
 with that for the period after 1930.
Thus, our choice of the reference period 1900--1976 is safe from this point of view.
We note that while this result is consistent with others \citep{sarychev09,carrasco13,aparicio14,willis16},
 it differs from that of \citep{cliver15,cliver16} who proposed a smooth parabolic ``learning curve'' of the RGO before 1920.

There is another interesting feature in Figure~\ref{fig:RGO-ratio} related to a bump during 1910--1930,
 when the RGO ratio is about 10--15\% higher than unity, suggesting that the RGO was counting more sunspot groups
 than other (normalized) observers.
Although it may not be excluded that it is not RGO showing higher values but other observers
 degrading in quality during that period (see an example of Wolfer below), the number of other observers during
 these years was five (Figure~\ref{fig:coverage}), and it is unlikely that they degraded simultaneously.
We note that this period was characterized by the change of the observers generations -- Wolfer, Quimby, Broger and Guillaume
 ceased their observations, while Luft, Brunner and later Madrid Observatory started.

The period after 1930 is characterized by the ratio around unity, implying a good consistency in the RGO data series.
Thus, the RGO series depicts a fair stability and is suitable to be the reference dataset, especially after 1890.

We have also tested the stability of the results vs. the exact choice of the reference dataset period.
While the main reconstruction is based on the RGO period 1900--1976, we have checked other periods as well.
The use of the full RGO dataset 1878--1976 as the reference period leads to a systematic decrease of the $S_s$ values
 by $\approx 5$ msd in comparison with the values shown in Table~\ref{Tab:observers} for the calibrated observers.
The final result in this case appears very close to the present one, with slightly lower $G-$values, within the error bars.
The use of the RGO dataset for the period 1913--1976, which was stable according to \citet{cliver16}, leads to a bit
 poorer statistic and a systematic increase of the $S_s$ values by 5--10 msd.
The final series based on this reference period yields slightly higher $G-$values but still in agreement with the main result
 within error bars.
We have also tested the effect of removing the ``bump'' period of 1913--1933 (discussed above) from the reference period.
It appears similar to the previous case, viz. an increase in the $S_s$ values by 5--10 msd and the final series
 consistent with the main result within error bars.
We have also checked that shrinking the reference period even further to 1933--1976 completely smears the result
 for two reasons.
First, the statistic is low, only four solar cycles.
But even more important is the fact that the cycles after the 1940s were very active, not being representative for the
 normal level of solar activity, which is the basic condition of the ADF method.
For instance, there was not a single month with ADF=0 during the period of 1933--1976.
This leads to a formally very strong offset of the obtained $S_s$ values being 25--40 msd greater than those in
 Table~\ref{Tab:observers} and consequently to unrealistically high $G-$values.
Accordingly, we conclude that the method is robust against the exact choice of the start of the reference period in a wide
 range, from 1878 till 1913, but the use of only high solar cycles leads to a violation of the basic assumption of the
 ADF method.

\subsubsection{Wolfer}

\begin{figure}
\includegraphics[width=\columnwidth,clip=]{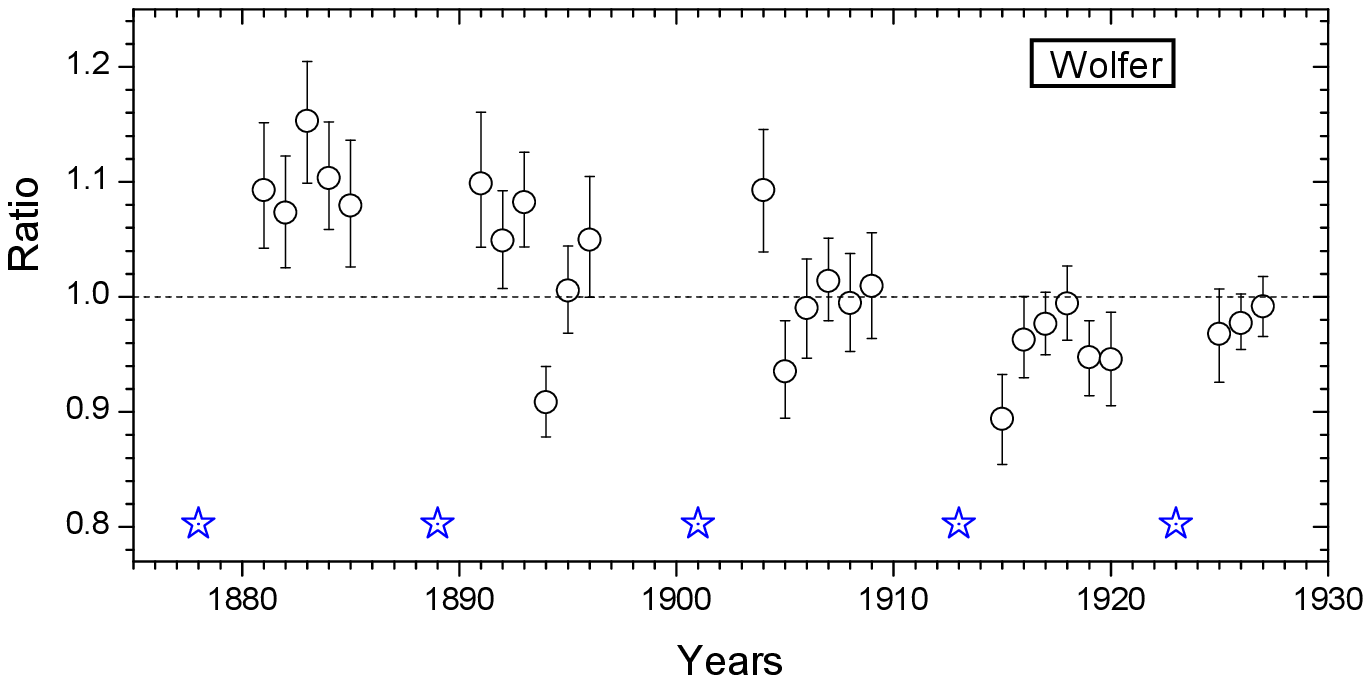}
\caption{The same as in Figure~\ref{fig:RGO-ratio} but for Wolfer.}
\label{fig:wolfer-ratio}
\end{figure}

We performed a similar analysis of the stability also for Wolfer, who is the reference observer for the ISN\_v2 series
 \citep{clette14} and the primary "backbone" observer for the S16 series.
The result is shown in Figure~\ref{fig:wolfer-ratio}.
One can see that there is a clear trend implying that the quality of Wolfer as an observer was slightly degrading in time
 so that he observed 10\% more groups than others in the 1880s but 5\% less than other in the 1910s.
Overall trend is 10--15\% during his scientific life.
Thus, we can conclude that the quality of Wolfer was slightly degrading through years within 10--15\%.
We have checked that this trend is not caused by the putative drift in the RGO data \citep{cliver16}
 by excluding the RGO dataset from the denominator of the ratio shown in Fig.~\ref{fig:wolfer-ratio}.
The trends remains qualitatively unaltered.

\subsubsection{Wolf}

\begin{figure}
\includegraphics[width=\columnwidth,clip=]{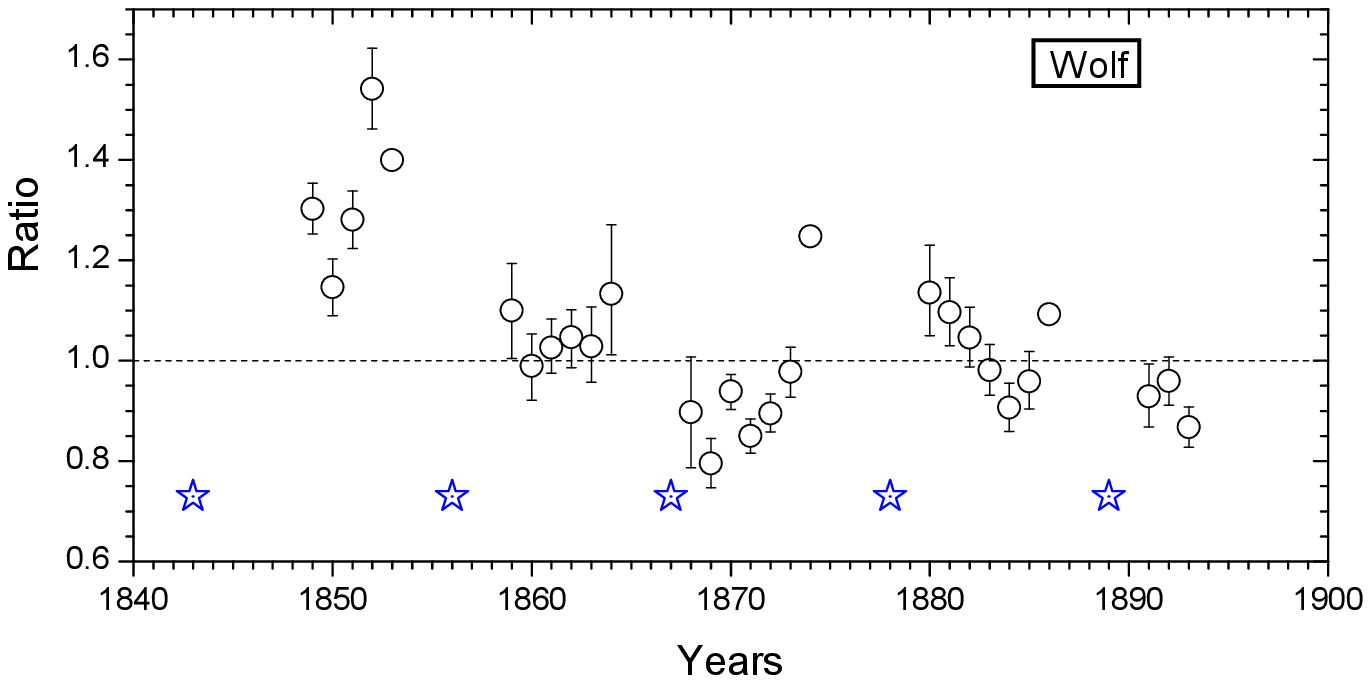}
\caption{The same as in Figure~\ref{fig:RGO-ratio} but for Wolf.}
\label{fig:wolf-ratio}
\end{figure}

Figure~\ref{fig:wolf-ratio} shows the ratios for Wolf, who is the reference observer for the WSN and ISN\_v1 series.
One can see that there is a clear enhancement in the beginning of the series around 1850, when Wolf counted
 about 30\% more groups relative to other observers.
This is likely related to the use of other (larger) telescope by Wolf.
However, since ca. 1860, his quality is around unity implying a fair stability within $\pm 20$\%.
Interestingly, for the period of overlap with Wolfer after 1880, the ratio for Wolf depicts a downward trend,
 which was interpreted by many as a sign of degradation of Wolf's eyesight \citep[e.g.,][]{clette14}.
However, it may be not the case since the ratio during the period 1880--1895 is fully consistent to
 that during the period of 1857--1875.

\subsubsection{Other observers}

We performed a similar analysis also for other observers as well (not shown) and found no specific features to be mentioned.
We note that this method of using the ratio works only if the number of overlapping observers is high enough.
Accordingly, when the number of regular observers is less than four, it becomes unclear.
Unfortunately, because of this, we cannot evaluate the stability of crucial observers before 1850.

This analysis suggests that for some, especially long-observing, observers an assumption on the stability
 of their observational quality may be not exactly valid.
However, this assumption makes a basis for all the existing sunspot series.
It will be a subject for a forthcoming work to assess this issue and to take it inot account.

\section{Discussion}

The final composite series of the number of sunspot groups constructed by the ADF method
 is shown in Figures~\ref{fig:monthly} (monthly values) and \ref{fig:annual}a (annual)
 along with the 68\% confidence intervals.

\subsection{Comparison with other series}
\label{sec:comp}
In Figure~\ref{fig:annual} we compare the results of this work with previously published annual
 values of the number of sunspot groups $G$: the original GSN series (divided by 12.08 to obtain the
 values of $G$) for the period 1610--1995 by \citet{hoyt98}, H98;
 the ``backbone'' $G-$series for 1610--2015 by \citet{svalgaard16}, S16;
 and the series, also based on ADF method, for the period 1749--1899 by U16.
It is important to say that the H98 series is calibrated to the reference RGO series
 using a $k-$factor method.
The normalization is direct for the period of 1874--1976 covered by the RGO data and includes
 a daisy-chain normalization outside this period.
The S16 series is based on the ``backbone'' method which uses key backbone observers,
 calibrated to the reference one.
The backbone observers were Staudacher, Schwabe, Wolfer and Koyama, who did not directly overlap
 with each other (see Table~\ref{Tab:observers})
 and thus can be linked together only via a multi-step daisy-chain procedure of linear normalization
 by means of $k-$factors.
As the reference observer, Wolfer was selected, and thus the S16 series is free of daisy-chain
 calibration only for the period 1880--1928, when direct Wolfer data is available.
For all other periods it includes a multi-step daisy-chain normalization.
The U16 series uses the RGO dataset for the period 1900--1976 as the reference,
 but presents data only before 1900.
Normalization is performed by the ADF method which is free of daisy chain.
The present result is also calibrated to the RGO dataset (1900--1976) using the ADF method.
For the period 1900--1976 we directly applied the ADF method but using the exact overlap of the
 observers with the RGO data, while a statistical comparison forms a basis for normalization outside this period.
This method is also free of daisy chaining.

The series are over-plotted in Figure~\ref{fig:annual}a, while panel b shows the ratio of individual series
 $G-$values to those of the present result for years with the annual number of sunspot groups
 not smaller than three.
Some specific periods can be identified for a detailed discussion.

After 1910, the present result is fully consistent with the H98 series, the ratio is around unity
 ($1.01\pm 0.04$).
This is understandable since both series are directly calibrated to the RGO dataset during this period,
 and the quality and quantity of observers was high in the 20th century.
Accordingly, the number of groups is most precisely defined for this period.
However, the S16 series is systematically lower by about 10\% ($0.90\pm 0.04$) in the 20th century.
This discrepancy is likely related to the normalization method of \citet{svalgaard16}, which
 uses Wolfer and Koyama as ``backbones'' for the 20th century.
Since these two observers did not overlap, \citet{svalgaard16} used cross-normalization including a multi-step
 daisy-chain procedure to reduce Koyama backbone data to the reference Wolfer conditions, that may
 introduce additional uncertainties.
We note that the data of Koyama agree with the result of this work (Figure~\ref{fig:all-obs}) within 5\% for the
 period 1947--1976 (overlap between Koyama and the RGO reference dataset).
Unfortunately, full information of the calibration for this backbone is not available from \citet{svalgaard16}
 to investigate this question in full detail.

For the period 1880--1900, the present result is in full agreement with the S16 series (the mean ratio over this period
 is $r=0.98\pm 0.05$),
 and we consider it as a good sign, since this was the period of Wolfer (the reference observer for the S16 series)
 observations, when no daisy-chain normalization was applied in the S16 series.
On the other hand, the H98 series is by 10\% too low ($r=0.89\pm 0.04$), probably related to the inhomogeneity
 in the RGO data series in its earlier part (see Section~\ref{sec:RGO}), as noted by \citet{clette14} and \citet{cliver16}.
The U16 series is slightly lower than the present one but consistent with the unity ($0.96\pm 0.05$).

The middle of the 19th century (1830--1870) is characterized by a great excess of the S16 series by about 30\%
 ($r=1.29\pm 0.08$).
Keeping in mind that the S16 series agrees with our reconstruction for the period of the Wolfer observations
 (see above), we may propose that this discrepancy is related to the calibration of the Schwabe backbone to Wolfer
 via Wolf in the S16 series.
As argued recently \citep{lockwood_SP3_2016,usoskin_ADF_16,usoskin_k_16}, the use of the linear $k-$factor as a conversion between
 Wolf and Wolfer data may lead to an overcorrection.
The H98 series is on average consistent with the present result ($r=0.96\pm 0.1$) but the ratio is inhomogeneous.
While it is around unity before 1848, it is systematically lower by 10\% ($r=0.89\pm 0.05$) after that.
This implies that the data by Wolf were likely under-corrected by \citet{hoyt98} by about 10\%.
The U16 series is insignificantly lower than the present result ($r=0.96\pm 0.07$), being generally consistent with it,
 with the discrepancies related to the slightly updated methodology used here.
This difference may be related to the different restrictions to the rare observations applied here and in U16
 (see Sect.~\ref{sec:calib}) and can serve as an estimate of the corresponding uncertainty.

For the period before the Dalton minimum, the S16 series is slightly higher than the present result
 ($r=1.1\pm 0.03$), but the ratio is inhomogeneous.
While the $G-$values of the S16 series are consistent with our data before 1760, they are
 about 7\% higher than that in the 1760--1780s.
This suggests that the normalization of Horrebow can be a reason for the discrepancy.
The H98 series, on the contrary, is systematically and significantly lower ($r=0.76\pm 0.03$)
 that the present result before the Dalton minimum, suggesting that it may be underestimated for that period.
It is important to note that both the S16 and H98 series, based on the daisy-chain calibration procedure
 dramatically loose quality before the Dalton minimum because of the lack of high-quality data
 at the turn of 18th and 19th centuries.
This makes it very difficult if even possible to make a `calibration bridge' across the Dalton minimum
 to relate the observers of the Staudacher era to those of the Schwabe era.
Anyway, the uncertainties of the daisy-chain $k-$factor calibration grow significantly before the Dalton minimum.
The U16 series is somewhat lower than the present one for the 1750--1790s ($0.93\pm 0.03$), because of the
 different ways to normalize the data of Staudacher (Section~\ref{sec:stau}) who was the key observer for that period.
Although the ADF method is free of daisy-chaining, uncertainties are also large (10--15\%) for the 18th century
 (see the shaded areas in Figure~\ref{fig:annual}b) because of the sparse data.
The present series agrees with the U16 one within these uncertainties although the latter tend to
 run systematically over the lower bound.

To summarize, the level of sunspot group activity yielded by the new series presented here
 lies between those for S16 and U16, but significantly higher (by 24\%) than that for the H98 series,
 in the second half of the 18th century before the Dalton minimum.
Due to large uncertainties for that period, all the series but H98 are marginally consistent with each other.
The new series is consistent with U16 and marginally with H98 ones during the 19th century but is
 significantly lower (by 25--30\%) than the S16 one, in particular refuting the high level of activity during the mid-19th century
 suggested in the S16 series.
However, the new series agrees with the S16 one during the 1880--1890s, viz. during the period of observations
 by Wolfer who is the reference observer for the S16 series.
In the 20th century, when the quality and quantity of observational data was high, the new series
 is fully consistent with the H98 series but is significantly (10\%) higher than the S16 one.

\subsection{Centennial variability}

Although the sunspot activity is dominated by the 11-year Schwabe cycle, centennial variability is also
 apparent in the time evolution of the composed series.
It is expressed in low (e.g., during the Dalton minimum around 1800) and high (middle of the 20th century) solar cycles.
There is no established way to define the centennial variability.
Sometimes decadal or cycle-averaged values are used to represent the centennial evolution in consistency with
 cosmogenic isotope data \citep{usoskin_LR_17}, or a linear trend over the envelope of solar cycles
 is considered \citep{clette14}.
Here we use the non-parametric Singular Spectrum Analysis (SSA) \citep{vautard92}, which decomposes
 a time series into several components with distinct temporal behaviors and is very
 convenient to disentangle long-term trends and quasi-periodic oscillations.
The method is based on the Karhunen-Loeve spectral decomposition theorem \citep{kittler73} and Man\'e-Takens
 embedded theorem \citep{mane81,takens81}.

\begin{figure}
\includegraphics[width=\columnwidth,clip=]{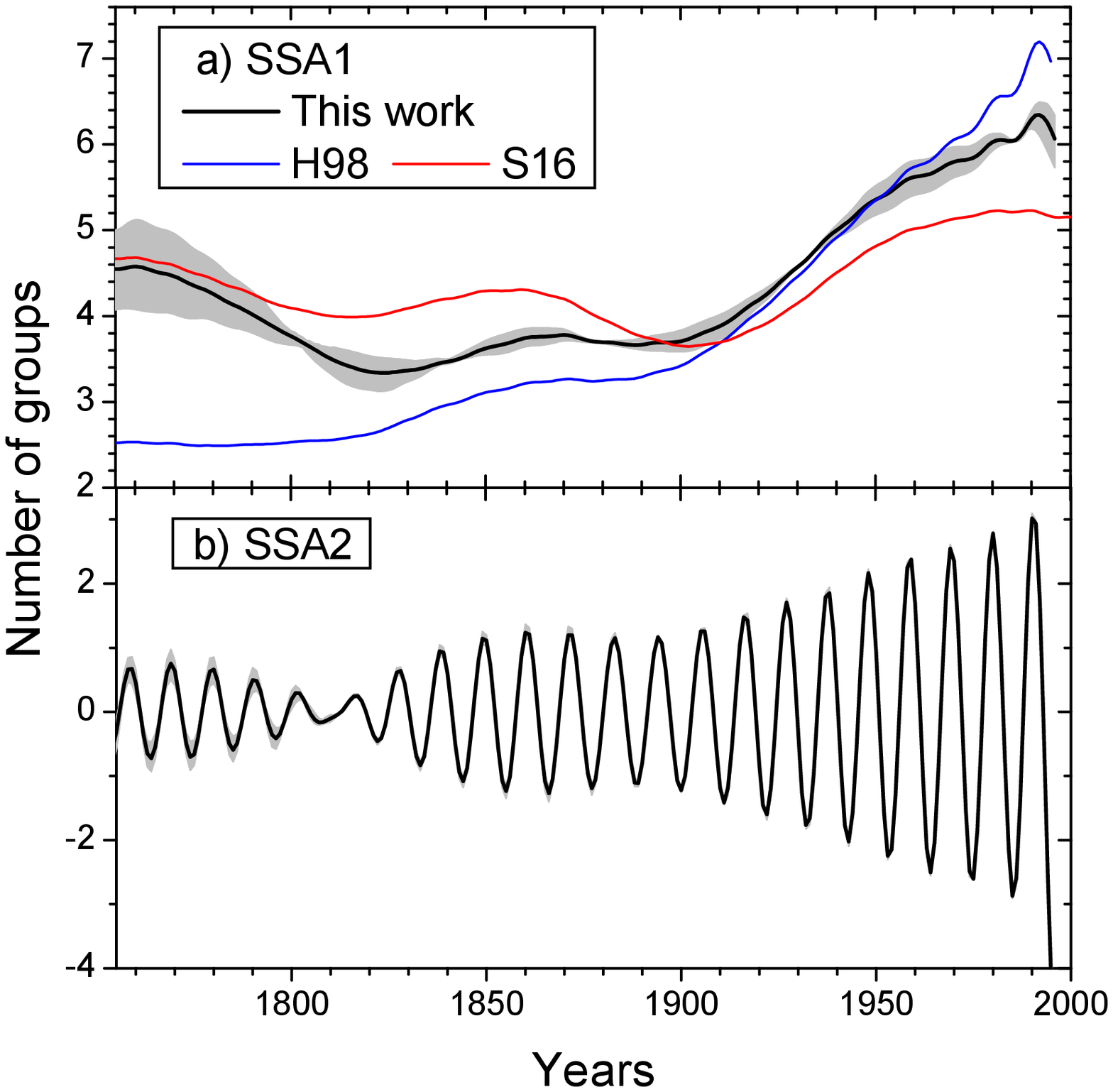}
\caption{First two components of the SSA analysis of the reconstructed series (panels A and B, respectively)
The black curves depict the mean while the shaded area the full range (corresponding to the time window of 40--80 years) of the
 SSA component values.
 The red and blue lines represent the first SSA component for the S16 and H98 series, respectively.}
\label{fig:SSA}
\end{figure}

The two first SSA components of the final composite series are shown in Figure~\ref{fig:SSA}.
We used the range of the time window for the SSA as 40--80 years, where the result is stable.
If the time window was too short, the 11-year cycle leaks into the first SSA component, while if it is too long,
 the pattern becomes smeared.
One can see that the time series is decomposed into a long-term trend or centennial variability (the first SSA component)
 and the 11-year cycle (the second component).
These two components represent 74\% of the overall variability of the series.

The series presented here depicts the relatively high activity in the mid-18th century ($G=4.5\pm0.5$)
 which decreases to 3.5--4 during the entire 19th century and then rises to around 6 in the second half of the 20th century.
This implies the significance of the Modern grand maximum of solar activity \citep{solanki_Nat_04}, so that the
 level of centennial sunspot activity in the second half of 20th century was a factor 1.33--1.77 higher
  than in the 18th and 19th centuries.

Figure~\ref{fig:SSA}a shows the first SSA components also for the H98 and S16 series (the U16 series is close to the
 one presented here and is not shown).
One can see different patterns of the centennial evolution (the primary SSA component) for these series.
The H98 series yields a monotonically increase of activity by a factor 2.5 between the mid-18th and late 20th centuries.
On the contrary, the S16 series suggests a roughly constant, slightly oscillating with about 100-yr period, level in the
 range of $G$ between four and five, without a clear grand maximum.
We note however, that the existence of the Modern grand maximum is independently confirmed by data from cosmogenic
 isotopes \citep[e.g.][]{abreu08,steinhilber12,inceoglu15,usoskin_AAL_14}.

\section{Conclusions}

A new revisited series of the numbers of sunspot groups is presented for the period 1749--1996, reconstructed by
 applying the active day fraction method to a revised database of sunspot group observations \citep{vaquero16}.
The new reconstruction agrees with the `classic' GSN series by \citet{hoyt98} in the 20th centuries
 but is systematically higher than that in the 18th century, suggesting a slightly higher than thought before
 solar activity in the mid-18th century.
On the other hand, the new series is systematically lower than that by \citet{svalgaard16} in the 18th
 and especially 19th centuries, implying that the latter overestimated the level of activity.

We have estimated the stability of some key solar observers.
The RGO dataset appears fairly stable against all other observers since the 1890s but is about 10\%
 too low before ca. 1885, as proposed earlier \citep{sarychev09,carrasco13,aparicio14,willis16,clette14}.
However, the conclusion by \citet{cliver16} that the RGO data are of uneven quality all the time
 before 1915, is not confirmed.
A declining trend of 10--15\% in the quality of Wolfer's observation is found between the 1880s and 1920s,
 suggesting that using him as the reference observer may lead to additional uncertainties.
On the other hand, Wolf (small telescope) appears fairly stable between the 1860s and 1890s, without
 an obvious trend.

The new reconstruction reflects the centennial variability of solar activity.
Using the SSA method, we decomposed different series into the primary centennial component (Figure~\ref{fig:SSA}a)
 and the secondary 11-year solar cycle.
The new series confirms the existence of the significant Modern grand maximum of solar activity in the second half
 of the 20th century, which appears a factor 1.33--1.77 higher than during the 18--19th centuries.
This is different from both the H98 series which shows a strong centennial trend with the growth of activity by a factor
 of 2.5 between the mid-18th and 20th centuries, and the S16 series which shows no centennial trend.
The existence of the Modern grand maximum is known independently also from cosmogenic isotope data
 \citep[e.g.][]{abreu08,steinhilber12,inceoglu15}.

The new series, available in Table 1 (annual values) and in CDS (monthly values), forms a basis for new studies of
 the solar variability and solar dynamo for the last 250 years.

\acknowledgement{
IGU and GAK acknowledge support by the Academy of Finland to the ReSoLVE Center of Excellence (project no. 272157).}

\begin{appendix}

\section{Annual number of sunspot groups}
\label{Sec:App_GMAG}

\begin{table*}
\caption{Annual numbers of sunspot groups: the mean, lower and upper 68\% quantiles, as shown in
 Figure~\ref{fig:annual}.
Missing values are denoted by -99.}
\label{Tab:annual}
\tiny
\begin{tabular}{cccc|cccc|cccc}
\hline
Year & $G$ & $G_{\rm low}$ & $G_{\rm up}$ & Year & $G$ & $G_{\rm low}$ & $G_{\rm up}$ & Year & $G$ & $G_{\rm low}$ & $G_{\rm up}$\\
\hline
1749	&	7.88	&	7.25	&	8.51	&	1800	&	2.25	&	2.58	&	2.92	&	1851	&	5.4	&	5.29	&	5.51	\\
1750	&	7.75	&	6.98	&	8.53	&	1801	&	4.53	&	4.64	&	4.75	&	1852	&	4.45	&	4.34	&	4.56	\\
1751	&	4.75	&	4.26	&	5.25	&	1802	&	3.65	&	3.82	&	3.99	&	1853	&	3.43	&	3.34	&	3.51	\\
1752	&	4.62	&	3.97	&	5.26	&	1803	&	2.38	&	2.52	&	2.65	&	1854	&	1.7	&	1.63	&	1.77	\\
1753	&	2.52	&	1.45	&	3.6	&	1804	&	2.37	&	2.52	&	2.68	&	1855	&	0.73	&	0.68	&	0.77	\\
1754	&	1.19	&	1.06	&	1.31	&	1805	&	2.62	&	2.80	&	2.97	&	1856	&	0.42	&	0.39	&	0.46	\\
1755	&	0.71	&	0.63	&	0.78	&	1806	&	1.30	&	1.55	&	1.79	&	1857	&	1.7	&	1.64	&	1.75	\\
1756	&	1.03	&	0.93	&	1.13	&	1807	&	2.65	&	3.25	&	3.82	&	1858	&	3.96	&	3.88	&	4.04	\\
1757	&	2.01	&	1.82	&	2.2	&	1808	&	1.09	&	1.55	&	2.01	&	1859	&	6.44	&	6.34	&	6.54	\\
1758	&	3.72	&	2.99	&	4.46	&	1809	&	0.77	&	0.98	&	1.19	&	1860	&	7.76	&	7.63	&	7.9	\\
1759	&	5.95	&	4.44	&	7.42	&	1810	&	0.10	&	0.40	&	0.70	&	1861	&	6.59	&	6.46	&	6.72	\\
1760	&	6.48	&	5.71	&	7.24	&	1811	&	-99	&	-99	&	-99	&	1862	&	4.93	&	4.83	&	5.02	\\
1761	&	7.61	&	7.29	&	7.94	&	1812	&	0.91	&	1.19	&	1.48	&	1863	&	3.94	&	3.83	&	4.04	\\
1762	&	6.22	&	5.86	&	6.59	&	1813	&	1.73	&	2.06	&	2.40	&	1864	&	3.65	&	3.56	&	3.75	\\
1763	&	4.64	&	4.24	&	5.04	&	1814	&	3.01	&	3.98	&	4.95	&	1865	&	2.33	&	2.26	&	2.4	\\
1764	&	3.01	&	2.65	&	3.36	&	1815	&	-99	&	-99	&	-99	&	1866	&	1.6	&	1.55	&	1.65	\\
1765	&	0.99	&	0.62	&	1.36	&	1816	&	3.84	&	4.32	&	4.79	&	1867	&	1.05	&	1	&	1.1	\\
1766	&	0.96	&	0.75	&	1.16	&	1817	&	3.76	&	4.07	&	4.36	&	1868	&	3.6	&	3.51	&	3.69	\\
1767	&	3.27	&	3.08	&	3.45	&	1818	&	2.12	&	2.30	&	2.47	&	1869	&	6.79	&	6.67	&	6.9	\\
1768	&	6.96	&	6.77	&	7.14	&	1819	&	2.17	&	2.38	&	2.60	&	1870	&	9.81	&	9.68	&	9.94	\\
1769	&	9.31	&	9.03	&	9.6	&	1820	&	1.73	&	1.94	&	2.14	&	1871	&	8.78	&	8.65	&	8.91	\\
1770	&	8.77	&	8.55	&	8.99	&	1821	&	3.06	&	3.65	&	4.24	&	1872	&	8.06	&	7.93	&	8.19	\\
1771	&	7.89	&	7.61	&	8.18	&	1822	&	2.45	&	3.06	&	3.67	&	1873	&	5.49	&	5.38	&	5.6	\\
1772	&	6.21	&	5.95	&	6.47	&	1823	&	1.12	&	1.38	&	1.65	&	1874	&	3.24	&	3.13	&	3.36	\\
1773	&	3.44	&	3.27	&	3.61	&	1824	&	1.65	&	1.79	&	1.94	&	1875	&	1.66	&	1.59	&	1.72	\\
1774	&	2.61	&	2.42	&	2.81	&	1825	&	1.9	&	1.82	&	1.98	&	1876	&	1.13	&	1.07	&	1.19	\\
1775	&	1.28	&	1.17	&	1.39	&	1826	&	2.56	&	2.51	&	2.61	&	1877	&	1.09	&	1.04	&	1.14	\\
1776	&	1.79	&	1.57	&	2.01	&	1827	&	3.67	&	3.62	&	3.72	&	1878	&	0.51	&	0.47	&	0.56	\\
1777	&	8.41	&	7.01	&	9.84	&	1828	&	4.89	&	4.83	&	4.95	&	1879	&	0.64	&	0.6	&	0.69	\\
1778	&	10.31	&	9.14	&	11.47	&	1829	&	4.87	&	4.81	&	4.94	&	1880	&	2.66	&	2.58	&	2.74	\\
1779	&	11.17	&	10.02	&	12.3	&	1830	&	5.37	&	5.31	&	5.43	&	1881	&	4.38	&	4.28	&	4.49	\\
1780	&	8.89	&	7.55	&	10.2	&	1831	&	3.64	&	3.59	&	3.7	&	1882	&	4.67	&	4.58	&	4.77	\\
1781	&	7.05	&	5.95	&	8.16	&	1832	&	2.11	&	2.06	&	2.15	&	1883	&	5.16	&	5.05	&	5.27	\\
1782	&	3	&	1.72	&	4.27	&	1833	&	0.6	&	0.57	&	0.63	&	1884	&	5.91	&	5.8	&	6.03	\\
1783	&	2.25	&	1.25	&	3.23	&	1834	&	0.91	&	0.88	&	0.94	&	1885	&	4.54	&	4.44	&	4.64	\\
1784	&	1.26	&	0.33	&	2.2	&	1835	&	3.77	&	3.7	&	3.85	&	1886	&	2.45	&	2.37	&	2.52	\\
1785	&	1.84	&	1.21	&	2.47	&	1836	&	7.21	&	7.13	&	7.29	&	1887	&	1.48	&	1.42	&	1.54	\\
1786	&	6.83	&	6	&	7.67	&	1837	&	7.85	&	7.77	&	7.93	&	1888	&	1.01	&	0.96	&	1.06	\\
1787	&	9.61	&	8.77	&	10.45	&	1838	&	5.98	&	5.92	&	6.04	&	1889	&	0.77	&	0.72	&	0.81	\\
1788	&	10.13	&	9.17	&	11.11	&	1839	&	5.04	&	4.98	&	5.1	&	1890	&	0.92	&	0.87	&	0.97	\\
1789	&	8.92	&	7.47	&	10.38	&	1840	&	3.87	&	3.82	&	3.91	&	1891	&	3.66	&	3.57	&	3.75	\\
1790	&	8.53	&	6.9	&	10.15	&	1841	&	2.33	&	2.28	&	2.37	&	1892	&	6.18	&	6.06	&	6.29	\\
1791	&	5.84	&	4.83	&	6.83	&	1842	&	1.62	&	1.58	&	1.66	&	1893	&	7.74	&	7.61	&	7.87	\\
1792	&	3.27	&	1.51	&	5.01	&	1843	&	0.76	&	0.72	&	0.79	&	1894	&	7.65	&	7.53	&	7.77	\\
1793	&	1.97	&	0.36	&	3.59	&	1844	&	1	&	0.97	&	1.03	&	1895	&	6.2	&	6.1	&	6.3	\\
1794	&	3.28	&	3.83	&	4.37	&	1845	&	2.49	&	2.44	&	2.53	&	1896	&	3.75	&	3.67	&	3.83	\\
1795	&	1.14	&	1.86	&	2.58	&	1846	&	3.52	&	3.47	&	3.57	&	1897	&	2.99	&	2.91	&	3.06	\\
1796	&	0.22	&	1.77	&	3.24	&	1847	&	4.75	&	4.67	&	4.84	&	1898	&	2.62	&	2.55	&	2.7	\\
1797	&	0.70	&	1.34	&	1.98	&	1848	&	6.78	&	6.69	&	6.88	&	1899	&	1.46	&	1.4	&	1.52	\\
1798	&	0.38	&	0.94	&	1.51	&	1849	&	7.18	&	7.05	&	7.3	&	1900	&	1.18	&	1.12	&	1.24	\\
1799	&	0.60	&	0.96	&	1.32	&	1850	&	5.21	&	5.1	&	5.32	&	1901	&	0.38	&	0.35	&	0.42	\\
\hline
\end{tabular}
\end{table*}

\begin{table*}
\caption{Continuation of Table~\ref{Tab:annual}}
\label{Tab:annual2}
\tiny
\begin{tabular}{cccc|cccc}
\hline
Year & $G$ & $G_{\rm low}$ & $G_{\rm up}$ & Year & $G$ & $G_{\rm low}$ & $G_{\rm up}$ \\
\hline
1901	&	0.38	&	0.35	&	0.42	&	1951	&	4.97	&	4.89	&	5.05	\\
1902	&	0.59	&	0.54	&	0.63	&	1952	&	2.58	&	2.53	&	2.64	\\
1903	&	2.31	&	2.24	&	2.37	&	1953	&	1.27	&	1.23	&	1.32	\\
1904	&	4.03	&	3.95	&	4.10	&	1954	&	0.60	&	0.56	&	0.63	\\
1905	&	5.34	&	5.25	&	5.44	&	1955	&	3.25	&	3.19	&	3.31	\\
1906	&	4.91	&	4.82	&	5.00	&	1956	&	10.13	&	10.04	&	10.23	\\
1907	&	5.48	&	5.39	&	5.57	&	1957	&	12.90	&	12.80	&	13.01	\\
1908	&	4.72	&	4.64	&	4.81	&	1958	&	13.36	&	13.26	&	13.47	\\
1909	&	4.13	&	4.05	&	4.21	&	1959	&	11.68	&	11.57	&	11.79	\\
1910	&	1.95	&	1.89	&	2.00	&	1960	&	8.36	&	8.27	&	8.45	\\
1911	&	0.78	&	0.74	&	0.81	&	1961	&	4.17	&	4.10	&	4.24	\\
1912	&	0.40	&	0.37	&	0.42	&	1962	&	2.88	&	2.82	&	2.93	\\
1913	&	0.20	&	0.18	&	0.23	&	1963	&	2.31	&	2.25	&	2.36	\\
1914	&	1.06	&	1.02	&	1.10	&	1964	&	1.16	&	1.12	&	1.20	\\
1915	&	4.24	&	4.16	&	4.32	&	1965	&	1.42	&	1.37	&	1.46	\\
1916	&	5.64	&	5.56	&	5.73	&	1966	&	3.73	&	3.66	&	3.80	\\
1917	&	8.53	&	8.43	&	8.62	&	1967	&	7.98	&	7.87	&	8.09	\\
1918	&	7.15	&	7.05	&	7.24	&	1968	&	8.00	&	7.91	&	8.10	\\
1919	&	5.65	&	5.57	&	5.74	&	1969	&	8.05	&	7.95	&	8.15	\\
1920	&	3.48	&	3.42	&	3.55	&	1970	&	8.75	&	8.65	&	8.86	\\
1921	&	2.41	&	2.36	&	2.47	&	1971	&	6.00	&	5.91	&	6.08	\\
1922	&	1.42	&	1.39	&	1.45	&	1972	&	5.99	&	5.90	&	6.08	\\
1923	&	0.74	&	0.71	&	0.76	&	1973	&	3.43	&	3.36	&	3.49	\\
1924	&	1.52	&	1.48	&	1.55	&	1974	&	3.02	&	2.95	&	3.08	\\
1925	&	4.23	&	4.17	&	4.29	&	1975	&	1.46	&	1.41	&	1.51	\\
1926	&	5.84	&	5.80	&	5.89	&	1976	&	1.36	&	1.31	&	1.40	\\
1927	&	6.28	&	6.22	&	6.33	&	1977	&	2.64	&	2.58	&	2.70	\\
1928	&	6.81	&	6.75	&	6.86	&	1978	&	8.66	&	8.53	&	8.79	\\
1929	&	6.21	&	6.15	&	6.27	&	1979	&	12.70	&	12.56	&	12.84	\\
1930	&	3.69	&	3.65	&	3.73	&	1980	&	10.56	&	10.45	&	10.67	\\
1931	&	2.15	&	2.12	&	2.18	&	1981	&	10.50	&	10.38	&	10.62	\\
1932	&	1.29	&	1.26	&	1.32	&	1982	&	9.13	&	9.01	&	9.24	\\
1933	&	0.62	&	0.60	&	0.65	&	1983	&	5.70	&	5.61	&	5.79	\\
1934	&	0.91	&	0.89	&	0.93	&	1984	&	3.71	&	3.64	&	3.77	\\
1935	&	3.67	&	3.61	&	3.72	&	1985	&	1.45	&	1.40	&	1.50	\\
1936	&	7.52	&	7.44	&	7.60	&	1986	&	1.09	&	1.05	&	1.13	\\
1937	&	10.09	&	9.99	&	10.18	&	1987	&	2.22	&	2.17	&	2.28	\\
1938	&	9.48	&	9.37	&	9.58	&	1988	&	6.79	&	6.69	&	6.89	\\
1939	&	7.75	&	7.69	&	7.82	&	1989	&	11.45	&	11.32	&	11.58	\\
1940	&	5.98	&	5.92	&	6.05	&	1990	&	11.76	&	11.63	&	11.89	\\
1941	&	4.22	&	4.16	&	4.28	&	1991	&	11.39	&	11.26	&	11.52	\\
1942	&	2.86	&	2.82	&	2.90	&	1992	&	7.84	&	7.74	&	7.94	\\
1943	&	1.52	&	1.48	&	1.56	&	1993	&	4.53	&	4.47	&	4.59	\\
1944	&	1.22	&	1.17	&	1.27	&	1994	&	3.09	&	3.04	&	3.14	\\
1945	&	3.45	&	3.38	&	3.52	&	1995	&	1.78	&	1.73	&	1.83	\\
1946	&	8.15	&	8.07	&	8.23	&	1996	&	1.11	&	1.06	&	1.15	\\
1947	&	11.36	&	11.26	&	11.47	&	--	&	--	&	--	&	--	\\
1948	&	10.83	&	10.71	&	10.95	&	--	&	--	&	--	&	--	\\
1949	&	10.58	&	10.48	&	10.68	&	--	&	--	&	--	&	--	\\
1950	&	6.35	&	6.27	&	6.43	&	--	&	--	&	--	&	--	\\
\hline
\end{tabular}
\end{table*}

\end{appendix}


\end{document}